# Simulating RISTRETTO: Proxima b detectability in reflected light

Maddalena Bugatti[1], Christophe Lovis[1], Nicolas Billot[1], Nicolas Blind[1], Baptiste Lavie[1], Martin Turbet[23], Bruno Chazelas[1], and Francesco Pepe[1]

[1] Département d'Astronomie, Université de Genève, Chemin Pegasi 51, CH-1290 Versoix, Switzerland
e-mail: maddalena.bugatti@unige.ch
[2] Laboratoire de Météorologie Dynamique, IPSL, CNRS, Sorbonne Université, 4 place Jussieu, F-75252 Paris Cedex 05, France
[3] Laboratoire d'astrophysique de Bordeaux, Univ. Bordeaux, CNRS, B18N, allée Geoffroy Saint-Hilaire, 33615 Pessac, France

**ABSTRACT**

*Context.* The characterization of exoplanet atmospheres is one of the key topics in modern astrophysics. To date, transmission spectroscopy has been the primary method used, but upcoming instruments, such as those on the European Extremely Large Telescope (ELT), will lay the foundation for advancing reflected-light spectroscopy. The main challenge in this area of research is the high contrast ratio between the planet and its star, especially for Earth-like planets, making it extremely difficult to isolate the planetary signal from that of its host star. RISTRETTO, a high-resolution integral-field spectrograph designed for ESO's VLT, aims to address these limitations through a combination of extreme AO, coronagraphy, and high-resolution spectroscopy.

*Aims.* The goal of this paper is to demonstrate the detectability of the temperate rocky planet Proxima b with RISTRETTO, using realistic end-to-end simulations and a specifically developed data analysis methodology.

*Methods.* We generated synthetic observations ensuring that the simulated spectra accurately reflect the complexities of real observational conditions. First, we created high-resolution star and planet spectra, selecting realistic observational epochs and conditions and incorporating the predicted performance of the AO and coronagraphic systems. Finally, we implemented noise and spectrograph effects through the Pyechelle spectrograph simulator. We then applied a state-of-the-art methodology to isolate the signal of the planet from the one of its host star and proceeded to fit several planetary models in order of increasing complexity. We also introduced a method to determine the sky orientation of the stellar spin axis, which constrains the orientation of the planetary orbit for aligned systems.

*Results.* Assuming an Earth-like atmosphere, our results show that RISTRETTO can detect Proxima b in reflected light in about 55 hours of observing time, offering the ability to characterize the planet orbital inclination, true mass, and broadband albedo. In addition, molecular absorption by $O_2$ and $H_2O$ can be detected in about 85 hours of observations.

*Conclusions.* These findings highlight the potential of RISTRETTO to significantly advance the field of exoplanetary science, by enabling reflected-light spectroscopy of a sample of nearby exoplanets ranging from gas giants to temperate rocky planets. This work sets the stage for detailed atmospheric characterization of Earth-like planets with next-generation AO-fed high-resolution spectrographs on extremely large telescopes, such as ELT-ANDES and ELT-PCS.

## 1. Introduction

In the last three decades, exoplanet research has undergone remarkable advances, with the discovery of thousands of planets exhibiting a wide range of masses, sizes, temperatures, and distances from their stars (Mayor & Queloz 1995; Borucki et al. 2010). However, detecting Earth-like exoplanets remains a significant challenge, particularly due to the low mass and radius ratios between rocky planets and their host stars. These typically translate into sub-m/s radial velocity signals and $10^{-4}$ photometric transit depths, which remain difficult to detect for month-to-year-long orbital periods, where the habitable zone is located (Fischer 2021; Pollacco et al. 2021). To address these challenges, recent efforts have focused on smaller stars, particularly M-dwarfs, whose reduced size and luminosity offer more favorable transit depths, more frequent transits, and increased transit probabilities for habitable-zone exoplanets (Dressing & Charbonneau 2013; Bonfils et al. 2013; Mulders et al. 2015; Gaidos et al. 2016).

While detecting exoplanets is crucial, spectroscopic characterization is necessary to explore their atmospheric properties and deepen our understanding of exoplanet formation and evolution (Madhusudhan 2019). Transmission spectroscopy, which involves analyzing starlight filtered through a planet atmosphere during transit, is one of the most commonly used techniques to probe exoplanets (Seager & Sasselov 2000). This approach reveals atmospheric features such as Rayleigh scattering (Lecavelier Des Etangs et al. 2008), clouds (Lendl et al. 2016), and the presence of various atomic and molecular species (Snellen et al. 2008; Snellen 2025). However, for smaller planets, multiple transits may need to be co-added to obtain sufficient signal-to-noise ratios (S/Ns), which might be prohibitive for long-period planets (Foreman-Mackey et al. 2016; Fatheddin & Sajadian 2024; Gialluca et al. 2021). The James Webb Space Telescope (JWST) offers unprecedented capabilities and versatility for studying the atmospheres of small exoplanets through transmission and emission spectroscopy. Examples of atmospheric characterization of warm to temperate rocky planets with JWST include the TRAPPIST-1 system (Zieba et al. 2023; Ducrot et al. 2023), or LHS 1140 (Cadieux et al. 2024). However, transit spectroscopy currently faces a major challenge that limits the ability to observe Earth-like exoplanets with JWST: stellar contamination from the transit light source (TLS) effect. This phenomenon arises from irregularities on the stellar surface, such as spots and faculae, which can significantly contaminate the measurements (Rackham et al. 2018; Lim et al. 2023).

For exoplanets that do not transit their host stars, thermal emission and reflected-light spectroscopy are the only means of characterization (Selsis et al. 2011; Hoeijmakers et al. 2018; Martins et al. 2015). The former approach analyzes thermal radiation emitted by the planet, providing insights into its temperature and atmospheric composition (Lewis et al. 2014), but it is particularly challenging for small temperate planets as it requires high-sensitivity mid-infrared observations. Reflected-light spectroscopy, on the other hand, involves analyzing starlight reflected off the planet atmosphere or surface, revealing details about its albedo and atmospheric composition and properties (Damiano & Hu 2021). Significant progress in reflection spectroscopy is expected from the combination of high-contrast imaging and high-resolution spectroscopy. This technique was initially proposed by Sparks & Ford (2002) and has since been refined in later studies (e.g. Snellen et al. (2015); Lovis et al. (2017)). Adaptive optics (AO) enables diffraction-limited observations with ground-based telescopes, providing sufficient angular resolution to separate short-period exoplanets from their host stars for nearby systems. Combined with AO, coronagraphy provides high-contrast capabilities at a few $\lambda/D$ from the star, an effective means of blocking contaminating starlight at the planet location; thus, it yields a contrast enhancement up to $10^3 - 10^4$ (Wang et al. 2017). A high spectral resolution then enables the separation of the planetary spectral signature from the one of the star based on their different Doppler shifts and spectral features. This combined approach is especially promising for characterizing Earth-like planets around M-dwarf stars, where the planet-to-star flux ratio can be as high as $10^{-7}$.

The innovative instrument known as high-Resolution Integral-field Spectrograph for the Tomography of Resolved Exoplanets Through Timely Observations (RISTRETTO) is an innovative instrument currently under development aiming at pioneering reflected-light spectroscopy of nearby exoplanets, based on the high-contrast, high-resolution approach (Lovis et al. 2024). It serves as a pathfinder for several major ground-based projects planned for extremely large telescopes (ELTs). Notable examples include ANDES and PCS on the European ELT (Palle et al. 2025; Marconi et al. 2022; Kasper et al. 2021) as well as IRIS and MICHI on the Thirty Meter Telescope (TMT) (Larkin et al. 2016; Packham et al. 2018). In parallel, space missions to characterize Earth-like atmospheres are also being developed. They include the Habitable Worlds Observatory (Feinberg et al. 2024), which aims to analyze the reflected light from terrestrial exoplanets, and the LIFE mission (Kammerer et al. 2022; Alei et al. 2024), designed to study their thermal emission with a mid-IR space-based interferometer.

The planet-to-star contrast in reflected starlight is given by (Horak 1950):

$$\frac{F_p(\lambda)}{F_s(\lambda)} = \left(\frac{R_p}{a}\right)^2 \cdot A_g(\lambda) \cdot g(\alpha), \quad (1)$$

where $F_p$ and $F_s$ are the planetary and stellar fluxes, $R_p$ is the planet radius, $a$ is the planet-to-star distance, $A_g(\lambda)$ is the geometric albedo, and $g(\alpha)$ is the phase function of the planet at phase angle, $\alpha$. The geometric albedo and phase function depend on the observation wavelength $\lambda$ and on the specific atmospheric and surface properties of the planet. The detectability of a planet reflected spectrum thus depends on planet size, orbital distance to the host star, atmospheric features, orbital phase, and orbital inclination. Angular separation on the sky between planet and star varies along the orbit, which can make the observation more difficult, and potentially unfeasible, if it falls below the inner working angle of the instrument. In addition, when a planet is fully illuminated (superior conjunction), it reflects more light but its relative Doppler shift with respect to the star decreases, complicating signal separation. Thus, a balance must be struck between maximizing reflected light and choosing an orbital phase that aids in distinguishing the planet signal from the one of the star based on their angular separation and relative Doppler shift.

Proxima Centauri b, located in the habitable zone of the M dwarf Proxima Centauri (Anglada-Escudé et al. 2016; Mascareño et al. 2020; Mascareño et al. 2025), presents an exciting opportunity for studying a potentially Earth-like exoplanet. Its proximity to Earth makes it an ideal target for reflected-light spectroscopy, within the reach of an instrument like RISTRETTO (Lovis et al. 2017), and future ELT instruments (Vaughan et al. 2024; Palle et al. 2025; Currie & Meadows 2025). In this paper, we study the detectability of Proxima b using RISTRETTO based on detailed end-to-end simulations and state-of-the-art data analysis. These simulations are crucial for understanding the spectrograph performance under various conditions, helping to optimize observing strategies and data analysis methods. By simulating RISTRETTO observations in detail, we can anticipate and address potential instrument limitations before actual observations are obtained, saving telescope time and ensuring robust data acquisition strategies and data analysis methods.

The problem at hand can be described as follows: the host star Proxima Centauri has an apparent $I$-band magnitude of $I = 7.4$. The stellar halo at the planetary location, as seen by RISTRETTO, typically corresponds to $I \simeq 17.4$ mag, namely, a factor of $10^4$ fainter. The planetary signal is expected to be another factor of $\sim 10^3$ fainter, corresponding to an apparent magnitude of $I \simeq 24.9$ mag. This highlights the need for exquisite suppression of stellar contamination and a robust data-analysis methodology to recover the planetary signal.

The paper is structured as follows. Section 2 presents the specifics of RISTRETTO and its data acquisition scheme. Section 3 details the simulations, from synthetic input spectra to recorded ones on the spectrograph detector. Section 4 discusses the data analysis procedures and data modeling. Section 5 gives planet detectability results. Section 6 presents a novel method to constrain the position angle of the stellar spin axis and, thus, the planet's position. Section 7 outlines our conclusions and future research directions.

## 2. RISTRETTO

### 2.1. Instrument description

RISTRETTO has been proposed as a visitor instrument for ESO's VLT. It is composed of two main parts. The front end, which includes an extreme AO (XAO) system that will provide diffraction-limited performance in the visible, along with a coronagraph to minimize the stellar brightness at the position of the planet. The back end consists of a seven-spaxel integral-field unit (IFU) feeding single mode fibers, and a high-resolution spectrograph covering the 620-840 nm wavelength range at $R =$ 140, 000. The RISTRETTO system overview is shown in Fig.1 (see Lovis et al. 2024; Chazelas et al. 2024). Below, we provide only a brief description of the instrument highlighting the properties that are needed to perform our simulations.

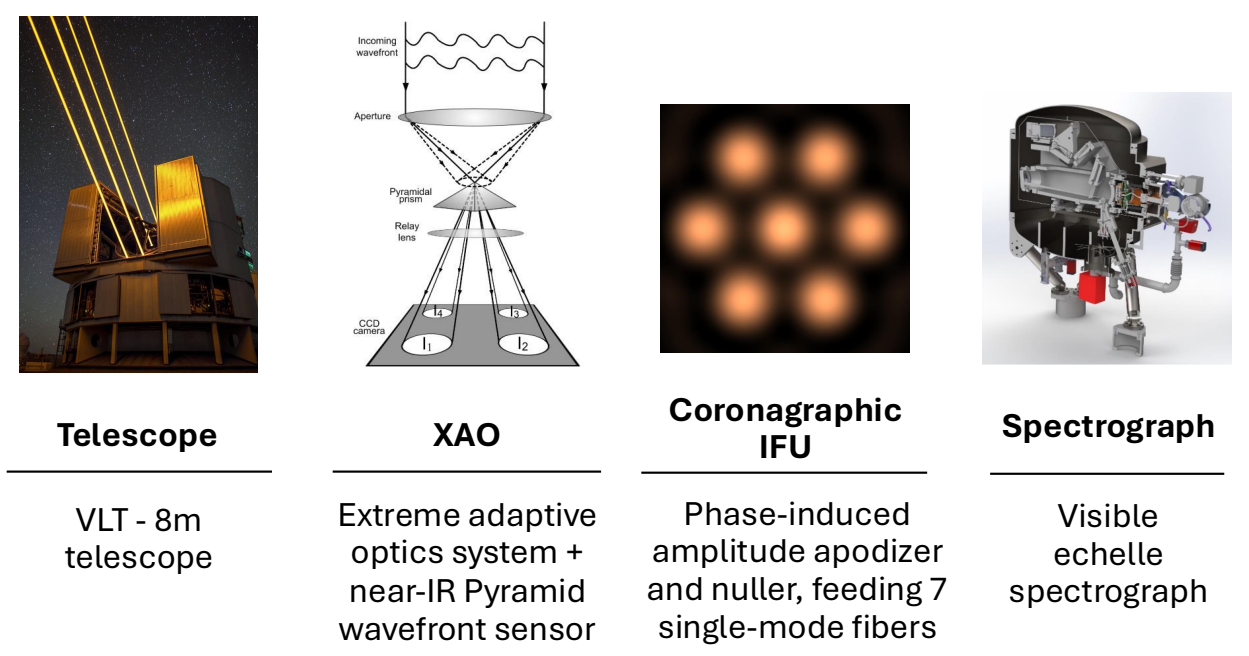

Fig. 1: RISTRETTO diagram.

#### 2.1.1. XAO front-end and coronagraphic IFU

The front-end design and performance are addressed in detail in Blind et al. (2024) and Blind et al. (2025). For our purposes, we summarize the front-end output as a set of IFU coupling curves, the derivation of which is described in Blind et al. (2024). The RISTRETTO IFU features seven hexagonal lenslets (spaxels); namely, a central one surrounded by six others, forming a ring at 2 $\lambda/D$ ($\sim$ 37 milliarcseconds at the VLT). Each lenslet directs light into a dedicated single-mode fiber, enabling diffraction-limited observations with a projection on the sky of approximately 37 milliarcseconds between the centers of adjacent spaxels. The coupling functions are obtained through full end-to-end simulations of the whole telescope+XAO+coronagraphic IFU system and describe how much light from a point source at any position within the field of view is coupled into each of the seven fibers.

#### 2.1.2. Spectrograph

We list below the main properties of the RISTRETTO spectrograph (Chazelas et al. 2024):

- Wavelength range: $\lambda_{min}$ = 620 nm, $\lambda_{max}$ = 840 nm;
- Spectral resolution: $R >$ 130, 000, goal 150,000;
- Echelle grating: MKS Newport R2 echelle grating with blaze angle of 63° and groove density of 23.2 lines/mm;
- Number of spectral orders: 32 orders, from the 92nd to the 124th;
- Pixel sampling: 2.5 pixels per resolution element on average in the main dispersion direction;
- Detector: CCD 231-84 from Teledyne-e2v with 4k x 4k pixels, RON $\sim$ 3 e-, dark current $\sim$ 0.5 e-/hour, QE $\sim$ 96% @ 650 nm (the official data sheet [1]);
- Wavelength calibration system: Uranium-neon lamp and Fabry-Perot interferometer

.

[1] https://www.e2v.com/resources/account/download-datasheet/1364\

#### 2.1.3. System transmission

The average total system transmission is about 7.7%, which includes the following contributions:

- Transmission of the atmosphere in the RISTRETTO science band at zenith: 97%;
- Telescope transmission with 3 aluminum-coated mirrors: 61%;
- Front-end transmission (optical components): 68%;
- Coupling into the coronagraphic IFU: depends on the AO performance, sky conditions, and on the object position within the IFU field of view. Planet coupling efficiency typically reaches between 30% and 50%, while stellar halo coupling remains below $2{\cdot}10^{-4}$ in the off-axis spaxels. A detailed discussion is provided in Sect.3.5;
- Fiber link efficiency: 89%;
- Average spectrograph efficiency (including detector): 44%.

These values are based on a combination of laboratory measurements, manufacturer data sheets, and prior experience with similar instruments.

### 2.2. Observing strategy

The goal of RISTRETTO is to detect an exoplanet signal in reflected light, but isolating the planet spectrum directly is not possible since the halo of the star always dominates the signal at the planet location. To overcome this, we envisage an observing strategy made of two types of exposures: a first spectrum is taken with both the planet and the stellar halo, followed by another with the star only. The second spectrum is used as a template to model the stellar halo in the first exposure, helping to isolate the planet spectrum. The first exposure will typically be 1 hour long to record the faint planetary signal while minimizing readout noise. In this exposure the host star will be centered in the central spaxel of RISTRETTO and the six off-axis spaxels will capture the stellar halo, with one of them also containing the exoplanet signal. During the long exposure a filter will be used in the central spaxel to prevent saturation on the detector.

Before and after each long exposure, a short exposure of typically a few minutes is taken with the host star centered on each of the six off-axis spaxels consecutively, using the tip-tilt mirror in the science channel to rapidly move the stellar PSF core across the IFU. This yields a high-S/N star-only spectrum in each IFU spaxel. This step is crucial for later modeling of the stellar halo spectrum, mitigating the effects of the different point spread function in the spectrograph focal plane for each off-axis spaxel. A visual diagram of the long+short exposures is shown in Fig.2. For the data analysis, we use $F_1$ to denote the long-exposure spectrum in the off-axis spaxel where the planet is located, while $F_2$ refers to the short exposure with the star-only spectrum injected in the same spaxel as the planet.

## 3. Simulations

In our simulations, we followed a step-by-step process to mimic real observations. We also aimed to ensure that the simulated spectra accurately reflect the complexities of real observational conditions:

3.1 Generating synthetic spectra: we crafted synthetic spectra for the host star, its surrounding halo, and the orbiting exoplanet;
3.2 Defining orbital parameters: for the simulation, we assumed some orbital parameters governing the trajectory of the target exoplanet in the plane of the sky;

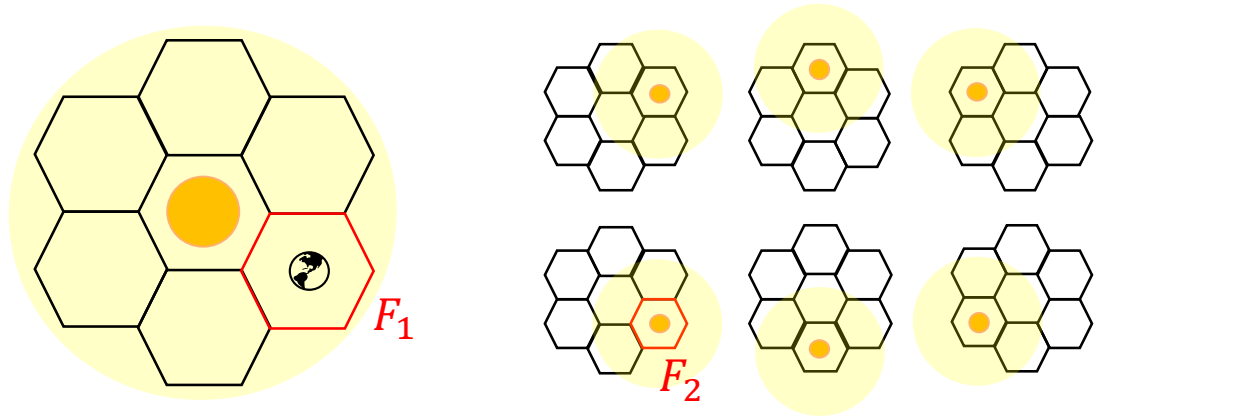

Fig. 2: Left panel: RISTRETTO long exposure with the IFU centered on the star and the planet in the off-axis spaxel. Right panel: Short exposure with the star illuminating all off-axis spaxels by performing a fast annular dithering pattern using the RISTRETTO tip-tilt mirror.

3.3 Selecting the epochs of observation: we established specific observing constraints and generate a realistic observational schedule;
3.4 Computing the radial velocities: we precisely computed the radial velocities for planet and star, and we apply the corresponding Doppler shift to each spectrum;
3.5 Incorporating the efficiency and IFU coupling functions: we included the system efficiency and coupling functions into the coronagraphic IFU for each spectrum, accounting for airmass and seeing conditions;
3.6 CCD properties: we incorporated realistic CCD parameters;
3.7 Generating 2D spectra: we used Pyechelle, a Python package, to generate 2D spectra based on the spectrograph optical design, incorporating the Point Spread Functions (PSF), photon and read-out noise, and realistic telluric absorption (Stürmer et al. 2018).

### 3.1. Generating synthetic spectra

To compute the input spectra, we used the data given in Table 1.

Specifically, to produce the star spectrum we used the Python package Expecto (Morris 2021), using as input the effective temperature, $T_{eff}$, of the star and its surface gravity $\log(g)$. Expecto uses the PHOENIX models (Husser et al. 2013) to create the spectrum emitted on the surface of the star in units of $erg/s/cm^2/cm$. Concerning the contribution from the planet Proxima b, we generated high-resolution (R=500,000) albedo spectra of the planet by coupling the outputs of a 3D Global Climate Model (Turbet et al. 2016) with the radiative transfer code PICASO (Batalha et al. 2019). We assumed an Earth-like atmosphere with a 1-bar, $N_2$-$O_2$-dominated atmosphere, with 400ppm of $CO_2$, and a surface fully covered with water. Water vapor and clouds are variable and calculated in a self-consistent way by the 3D model. Molecular absorptions from all four molecules ($N_2$, $O_2$, $CO_2$, and $H_2O$) were taken into account in our radiative transfer calculations, as well as Rayleigh and Mie scattering. For all the simulations, we used a reflected spectrum computed at an orbital phase angle of 90 degrees. We call the synthetic star spectrum $F_s(\lambda)$, the synthetic planet spectrum $F_p(\lambda)$, and the ratio between the two $F_p(\lambda)/F_s(\lambda)$. All of them are shown in Figure 3.

### 3.2. Defining orbital parameters

When detecting an exoplanet using the radial velocity method, such as Proxima b, four of the six orbital parameters can be determined (they are provided in Table 1), while the inclination angle ($i$) and the longitude of the ascending node ($\Omega$) of the orbit

Table 1: Proxima and Proxima b properties.

| Proxima Centauri | Value | ref |
|---|---|---|
| Right Ascension | 14 : 29 : 42.946132 | 1 |
| Declination | −62 : 40 : 46.16468 | 1 |
| Distance from Earth | $d = 1.3012 \pm 0.0003$ pc | 1 |
| Mass | $M_s = 0.1221 \pm 0.0022\ M_\odot$ | 2 |
| Radius | $R_\star = 0.141 \pm 0.021\ R_\odot$ | 3 |
| I-band magnitude | $I = 7.4$ | 7 |
| Rotational period | $P_{rot} = 83.1 \pm 1.4$ d | 4 |
| PHOENIX Model | | |
| | $T_{eff} = 3000$ K | 5 |
| | $\log(g) = 5$ | 6 |
| | $[Fe/H] = 0$ | 6 |
| | $[\alpha/M] = 0$ | 6 |
| **Proxima b** | **Value** | |
| $T_0$ −2450000 d | $T_0 = 10548.59 \pm 0.12$ d | 4 |
| Minimum Mass | $M_p \sin i = 1.055 \pm 0.055\ M_\oplus$ | 4 |
| Orbital Period [P] | $P = 11.18465 \pm 0.00053$ d | 4 |
| Semi-Major Axis | $a = 0.04848 \pm 0.00029$ AU | 4 |
| Eccentricity | $e = 0$ (fixed) | 4 |
| RV amplitude | $K_s = 1.226 \pm 0.062$ m/s | 4 |

**Notes.** References: 1-Gaia Collaboration et al. (2021), 2- Mann et al. (2015), 3- Boyajian et al. (2012), 4- Mascareño et al. (2025) ($T_0$ is the time of inferior conjunction), 5- Turbet et al. (2016), 6- Husser et al. (2013), 7-Ribas et al. (2017).

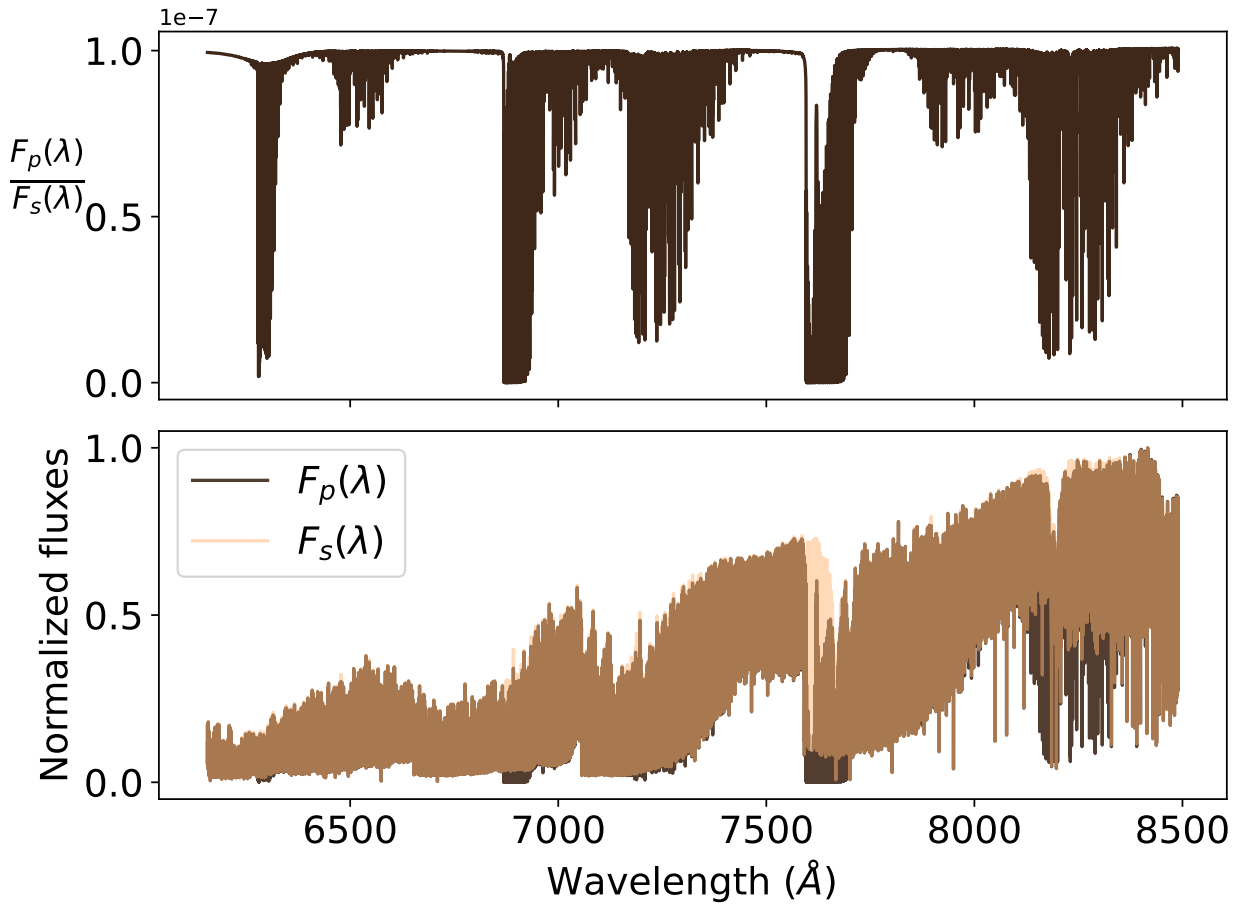

Fig. 3: Upper plot: Albedo spectrum $F_p/F_s$ of Proxima b for an Earth-like atmosphere. Lower plot: Normalized planetary ($F_p$) and stellar ($F_s$) spectra. $F_p$ is generated at an orbital phase of 90°. All spectra have a resolution of $\sim 500,000$.

plane (Hatzes 2016) remain unknown. Knowing these parameters is essential for accurately determining the orbit of the exoplanet in the plane of the sky as observed from Earth. Possible trajectories for Proxima b, as seen from Earth, are shown in Figure 4.

Specifically, knowing the ascending node longitude ($\Omega$) determines the exact position angle where the exoplanet will appear at its maximum elongation from the host star. Meanwhile, the inclination angle ($i$) is necessary to accurately calculate the orbital radial velocity of the exoplanet. In our simulations, we assume we can determine the longitude of the ascending node - as shown in Section 6. If $\Omega$ cannot be determined beforehand, RISTRETTO will mitigate this uncertainty by taking two consecutive exposures, each rotated 30 degrees relative to the other. This strategy provides more complete coverage of the stellar surroundings, increasing the chances of exoplanet detection (as illustrated in Figure 5) and compensates for the near-zero transmission between adjacent spaxels.

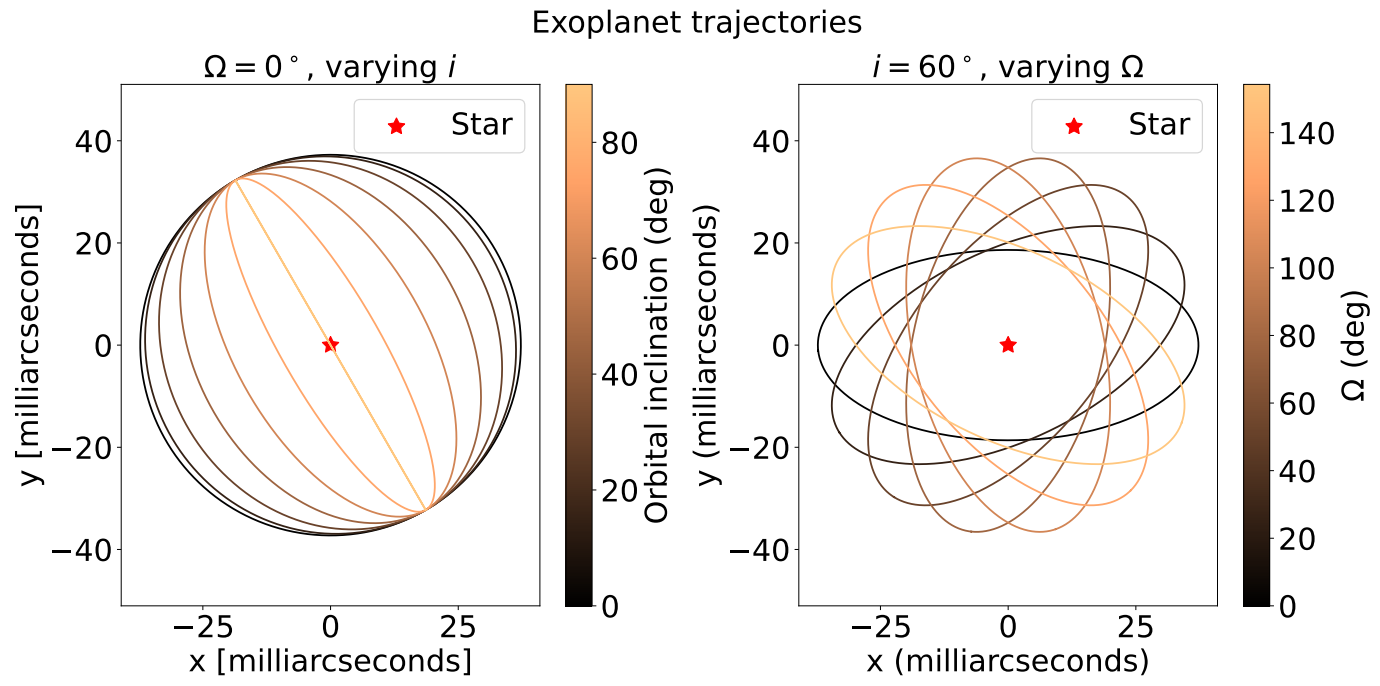

Fig. 4: Proxima b's possible trajectories as seen from Earth with one of the two unknown orbital elements ($i$ and $\Omega$) fixed.

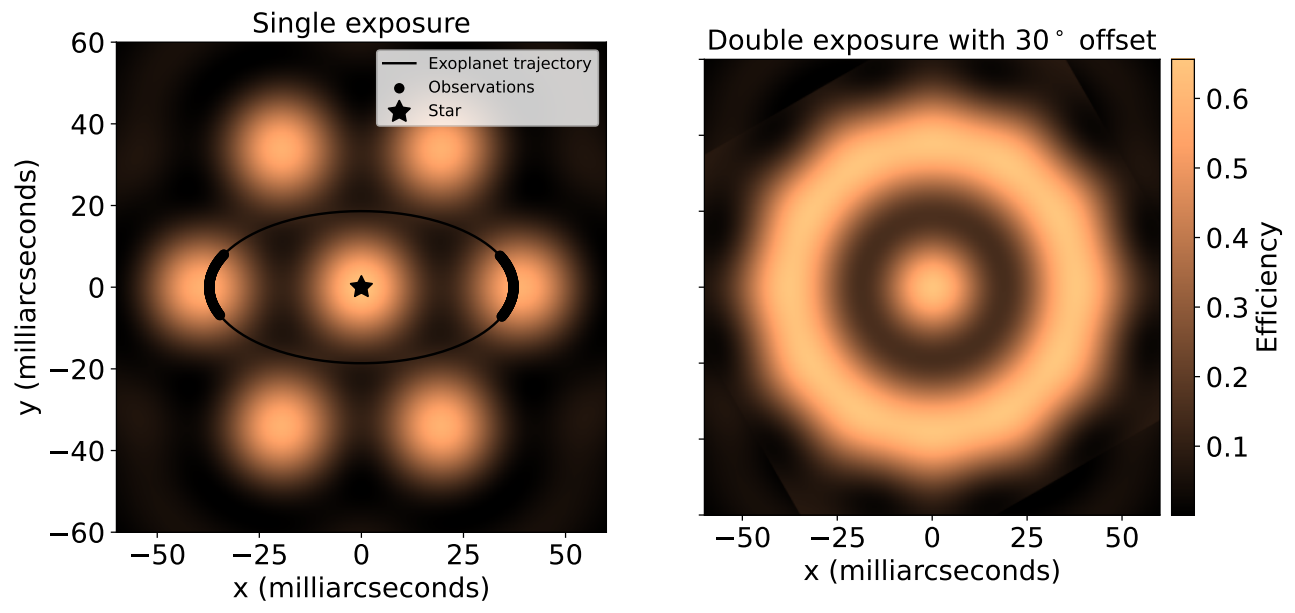

Fig. 5: Left panel: IFU transmission map into the 7 fibers for single orientation, with the trajectory of Proxima b overlaid on top. Right panel: IFU transmission combining 2 exposures with a 30° offset. For both plots, we assumed median seeing (0.76") and elevation (43°) conditions, and $\lambda$ = 840 nm.

For our simulations, we assumed two hypothetical values for the unknown parameters:

- The inclination of the orbit $i = 60°$;
- The longitude of the ascending node $\Omega = 0°$.

The choice for $\Omega$ was random (we chose $\Omega = 0$ for simplicity), while for the orbital inclination, we considered the probability density function based on a random distribution of orbital orientations in space. Specifically, the probability density function of $i$ is $P(i) = sin(i)$, where $i$ ranges from 0° (face-on orientation) to 90° (edge-on orientation). At an inclination of 60°, there is an equal probability of the inclination being higher or lower than this angle. The orbit of Proxima b on the sky can be seen in the left plot of Figure 5. It was computed by converting orbital elements of Table 1 to Cartesian coordinates (in milliarcseconds) in the plane of the sky.

### *3.3. Selecting the epochs of observation*

To make our simulation even more realistic, we planned specific dates and times for observing Proxima b, as observed from VLT, Paranal observatory, Chile (European Southern Observatory 2024), where RISTRETTO is planned to be used as a visitor instrument. To select the observation epochs, we applied several constraints to ensure optimal conditions:

- We limited observations to times when the altitude of the Sun is below −12 degrees, guaranteeing nautical night;
- We constrained the airmass to a maximum of 1.7, over the whole exposure time of any exposure. The choice for this limit is better explained in Section 5.1;
- The seeing must be better than the 75% percentile distribution of the local seeing in Paranal, corresponding to a seeing better than 0.97". This number will be justified in Section 5.1;
- We set a maximum allowed distance of 9 milliarcseconds between the planet and the center of any off-axis spaxel. This threshold corresponds to the separation at which the coupling efficiency of the off-axis spaxel is roughly half that at its center. This calculation assumes an inclination of $i = 60°$; at lower inclinations, the planet would move away from the center of the off-axis spaxel more rapidly as it follows its orbit. However, lower inclinations also correspond to a larger true planetary mass, strengthening the signal;
- We required a minimum of four consecutive hours of observation per night. Nights with fewer than four valid hours were excluded, as they would introduce inefficiencies in time and effort for a minimal observation duration.

Additionally, we assumed 1 hour of exposure time for the long exposure where the star is centered in the central spaxel and 2-minute short exposures where the star is centered on the off-axis spaxels. We considered that the short exposures would be taken immediately before and after each long exposure. Accounting for all the previously listed conditions, we found that in an observing season (year 2028), the total number of available long exposures is 241, while the number of available short exposures is 276.

Figure 6 shows the airmass and elevation of Proxima Centauri over the observational epochs (an elevation of 90° corresponds to Zenith) at the midpoint of each 1-hour long exposure. Moreover, the black points of Figure 5 represent the planet position on top of the planetary orbit during the long exposures.

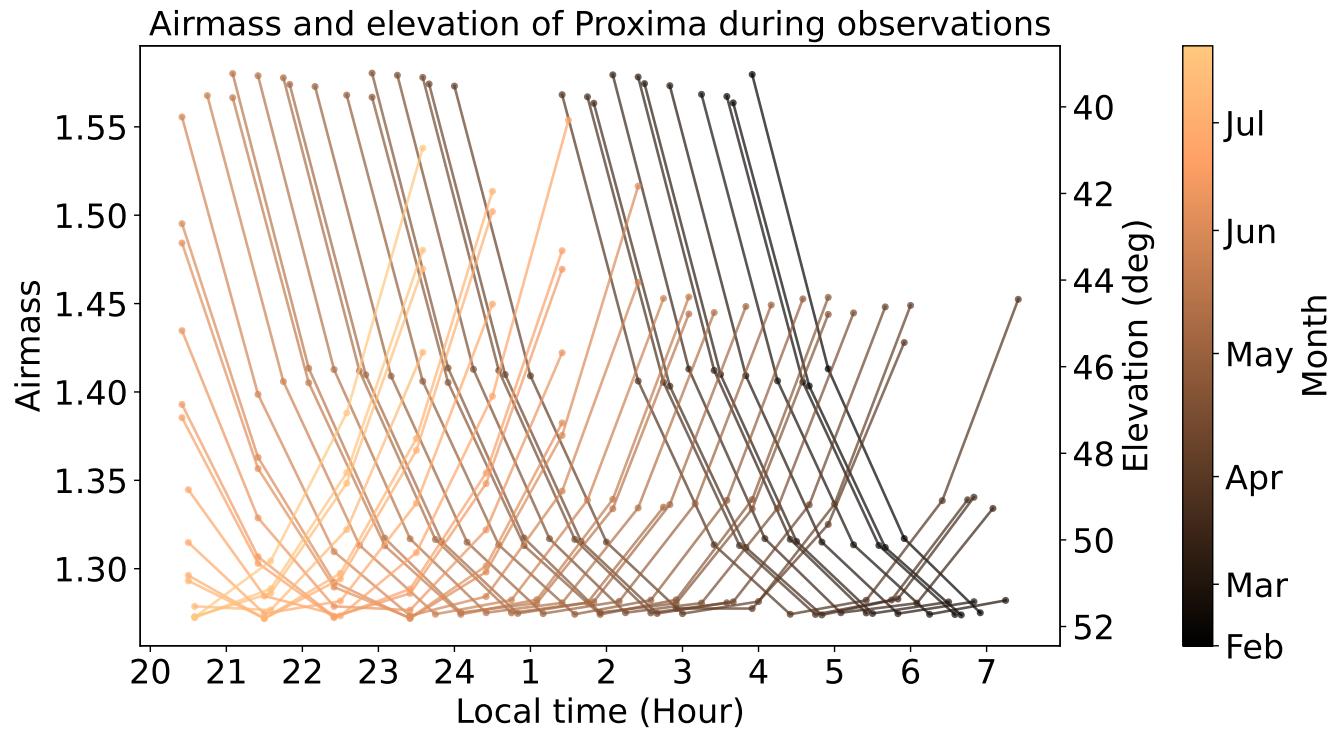

Fig. 6: Airmass and elevation of Proxima Centauri during the selected observational epochs - for the year 2028

In summary, there are 241 possible long exposures per year (under our observational constraints), for which Proxima b is suitably located.

### 3.4. Computing the radial velocities

We calculated the radial velocities of both the planet and the star throughout their orbits to adjust the Doppler shift in the input spectra for each exposure. We considered the following three contributions to the Doppler shift:

1. Systemic radial velocity, which is the radial velocity of the barycenter of the star + planet(s) exosystem, measured in the solar system barycentric rest frame. It has a fixed value and for the Proxima system is equal to −21.7 km/s (Lovis et al. 2017);
2. Orbital radial velocity, which refers to the radial velocity component of both star and planet caused by the orbital motion around the center of mass of the system.
   Using the data reported in Table 1 and $i = 60°$, we computed the orbital radial velocities for the star and the planet, along with the star-planet separation (shown in Figure 7). In black we give the observational epochs, which are close to the maximum elongation of the planet.

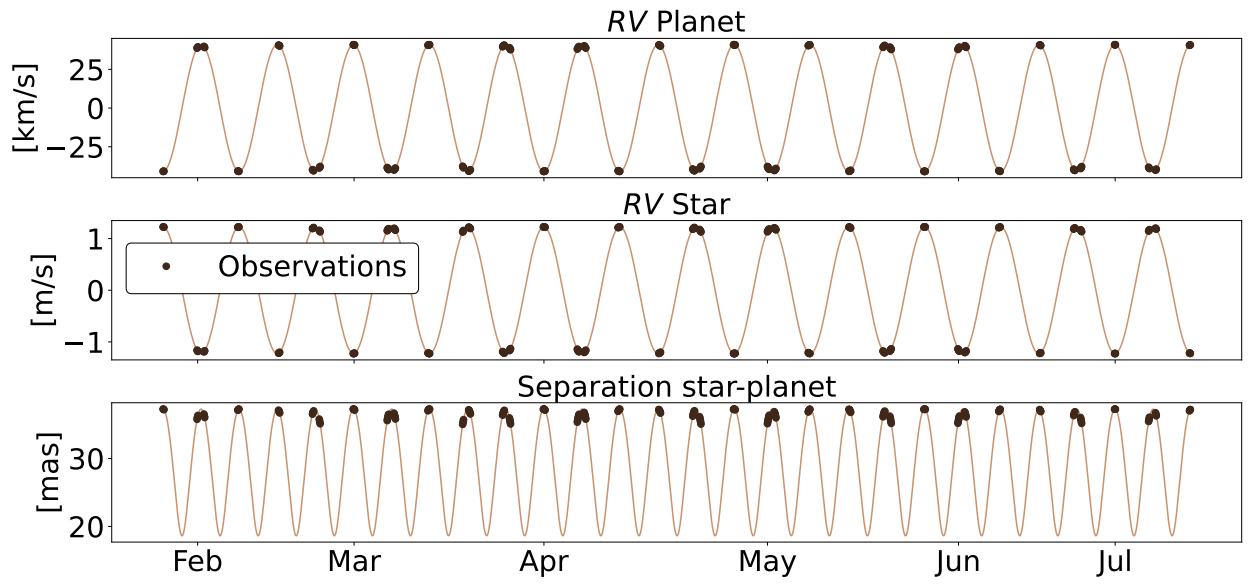

Fig. 7: Orbital radial velocity of star and planet and separation between the two bodies. The black points represent the selected observational epochs.

3. Barycentric correction - It refers to the velocity of the observer relative to the barycenter of the Solar System. Barycentric radial velocity includes the motions of the observer, such as the motion of the Earth around the Sun and the rotation of the Earth. This velocity is the same for both Proxima and Proxima b and depends on the date and time of the observations.
   We computed this latter radial velocity using the python package Barycorrpy (Kanodia & Wright 2018). In Appendix A, we give the Barycentric radial velocities during the observational epochs (2028).

### 3.5. Incorporating the efficiency and IFU coupling functions

As total system efficiency, we multiplied the individual efficiencies specified in Section 2.1. To estimate the IFU coupling, we used the results from extensive XAO simulations of Blind et al. (2024), which take into account the performance of both the coronagraphic IFU and the AO system (Restori et al. 2024; Blind et al. 2024). The simulations were conducted under various seeing and elevation conditions, as well as for different positions of the observed object on the IFU. For our analysis of Proxima b, we use the following combination of parameters: 5 wavelengths $[620, 675, 730, 785, 840]$ nm, 4 elevations $[38°, 43°, 48°, 52°]$, 4 seeing conditions at zenith $[0.52, 0.62, 0.76, 0.97]$ arcseconds (corresponding to the $10^{th}, 25^{th}, 50^{th}, 75^{th}$ percentile of the distributions at Paranal), and various positions on the IFU of the observed object (ranging from −60 to +60 milliarcseconds along both the $x$ and $y$ axes, with intervals of 2 milliarcseconds), applied to the 7 fibers. The choice not to include worse seeing conditions is explained in Section 5.1.

After selecting observational epochs, the elevation was determined for each epoch. Subsequently, seeing values were randomly drawn from the statistical distribution measured at Paranal (Section 5.1). For each observation, we identified the coupling maps with elevation and seeing values closest to the selected conditions. In computing the coupling maps the seeing values at zenith are converted to actual seeing at a given elevation using standard formulas (Roddier 1981). To account for intermediate positions corresponding to our specific observation points, we performed an interpolation across the five wavelengths and the $60 \times 60$ grid of $x$ and $y$ coordinates, measured in milliarcseconds.

Figure 8 shows an example of the coupling functions into all the fibers, for median conditions of seeing (0.76") and elevation (43°), as a function of wavelength. In the left panel an object is placed at the center of each spaxel, in the right panel the object is placed at the center of the central spaxel, which corresponds to fiber 0. The imbalance between the off-axis coupling functions on the bottom right plot of Figure 8 is primarily caused by the limited duration of the XAO simulations and is expected to diminish with longer runs. However, the simulations of the coronagraphic IFU do not account for residual atmospheric dispersion correction (ADC) correction (potentially at the level of 2 mas) which could further amplify the observed variations. Moreover, in practice, some differences will persist on the bench due to IFU imperfections or system asymmetries.

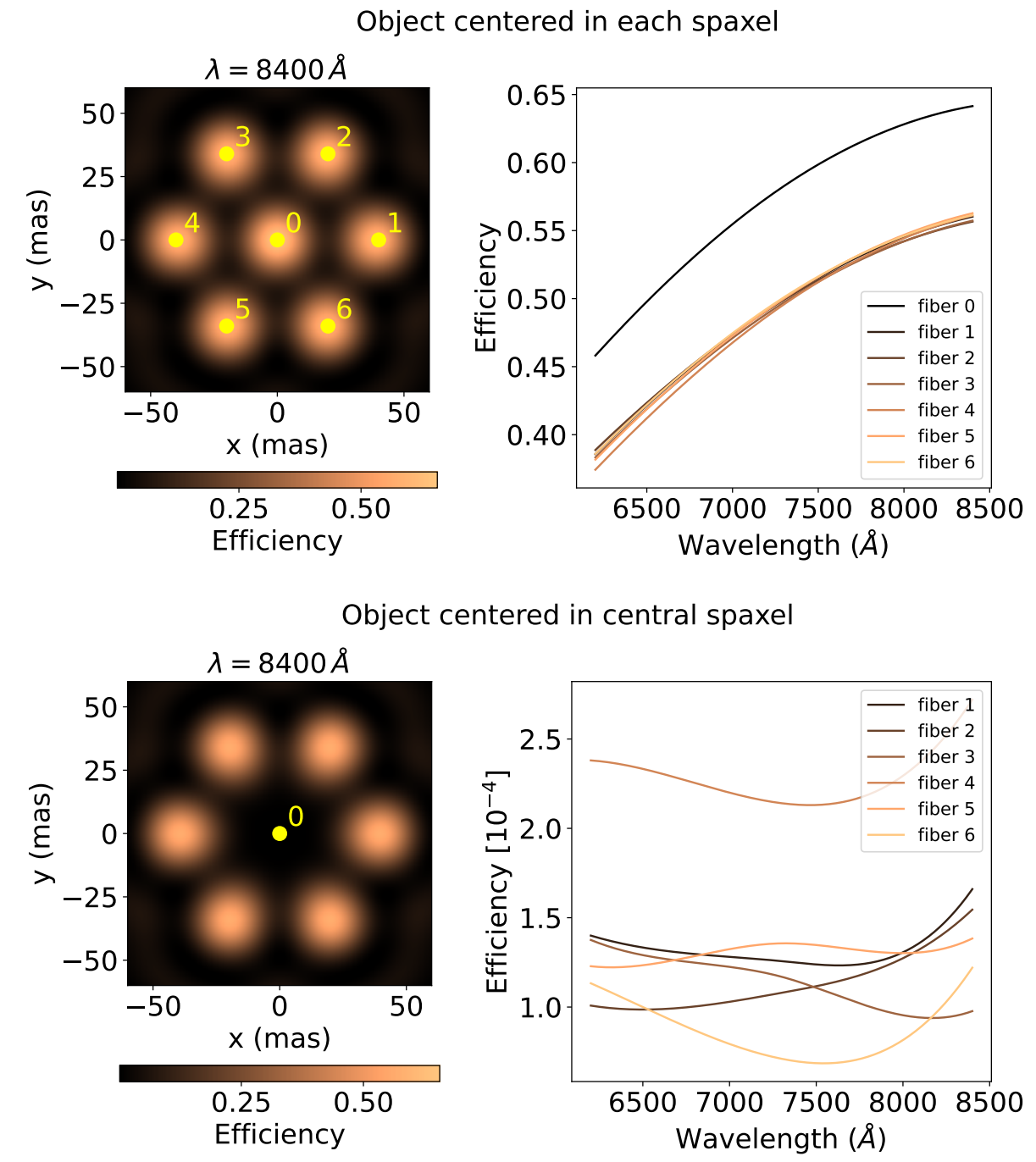

Fig. 8: Top plot: IFU coupling functions for the 7 fibers, when centering an object on each of them ($\rho_i^s(\boldsymbol{r}_{\rm off})$ of Section 4.2). Bottom plot: IFU coupling functions for the 6 off-axis fibers, when centering an object in the central spaxel-fiber 0 ($\rho_i^s(\boldsymbol{r}_{\rm on})$ of Section 4.2). We considered median conditions of seeing (0.76") and elevation (43°).

The coupling functions used in this paper were derived from simulations assuming realistic wavefront aberrations for the coronagraphic IFU (Blind et al. 2024). Additionally, knowing the longitude of the ascending node would enable us to observe the planet using two opposite fibers of our choice, ideally those with the lowest coupling for the stellar halo, namely fibers 2,3,6 by looking at the right panel of Figure 8. For this simulation, we chose to use fiber 6 and 3, which provide an average suppression factor of $1.1 \cdot 10^{-4}$ and $0.86 \cdot 10^{-4}$, respectively. These values are consistent with the technical specifications for the XAO and IFU systems of RISTRETTO.

### *3.6. CCD properties*

The model of the CCD in PyEchelle includes only two input parameters that we can set: the read-out noise and the bias level. The photon noise is generated automatically. To have an estimation of the read-out noise, we looked at the RISTRETTO CCD data sheet and decided to set it equal to 3 electrons. We arbitrarily set the bias level to 250 electrons. Dark current is not explicitly included in our simulation. According to the CCD data sheet and expected operational temperature, it amounts to about 0.5 electron/pixel/hour. Dark current noise is thus negligible compared to read-out noise.

### *3.7. Generating 2D spectra*

To simulate RISTRETTO spectra, we utilized the Python package PyEchelle (Stürmer et al. 2018), specifically designed for generating realistic 2D cross-dispersed echelle spectra. It operates on the principle that any spectrograph can be represented using wavelength-dependent transformation matrices and point spread functions to describe its optics. Transformation matrices transform one vector into another through matrix multiplication, preserving points, straight lines, and planes, though not necessarily distances or angles. In PyEchelle, these matrices and point spread functions are derived from the RISTRETTO model in ZEMAX, an optical modeling software (Zemax, LLC 2023). As the geometric transformations associated with the matrices vary smoothly across an echelle order, they are interpolated in PyEchelle for any intermediate wavelength. The same applies to the wavelength-dependent point spread functions, which slowly vary across an echelle order, allowing for interpolation at intermediate wavelengths (Stürmer et al. 2018). The matrix transformation model from RISTRETTO includes 10 wavelengths per order to construct the transformation matrices, and a point spread function (PSF) of 128x128 pixels for each wavelength. The scheme of Figure 9 shows the workflow of Pyechelle. PyEchelle generates random white noise to simulate read-out and photon noise in each pixel. Moreover, it also includes the possibility to simulate atmospheric telluric absorption according to realistic observing conditions and site - through the ESO Skycalc model (Noll et al. 2012).

Figure 10 displays a simulated 2D raw frame generated with PyEchelle. Each spectral order contains seven spectra, each of them representing one spaxel.

The wavelength solution varies notably across spaxels within the same order due to physical constraints at the entrance slit of the spectrograph. To accommodate all the desired orders on the CCD, a limit is imposed on the vertical distance between the centers of the fiber cores. The cladding of the fibers would prevent us from achieving this required vertical spacing if the fibers were stacked directly on top of each other. Therefore, the fibers were horizontally shifted across the entrance aperture, which leads to a significant variation in the wavelength solution from one

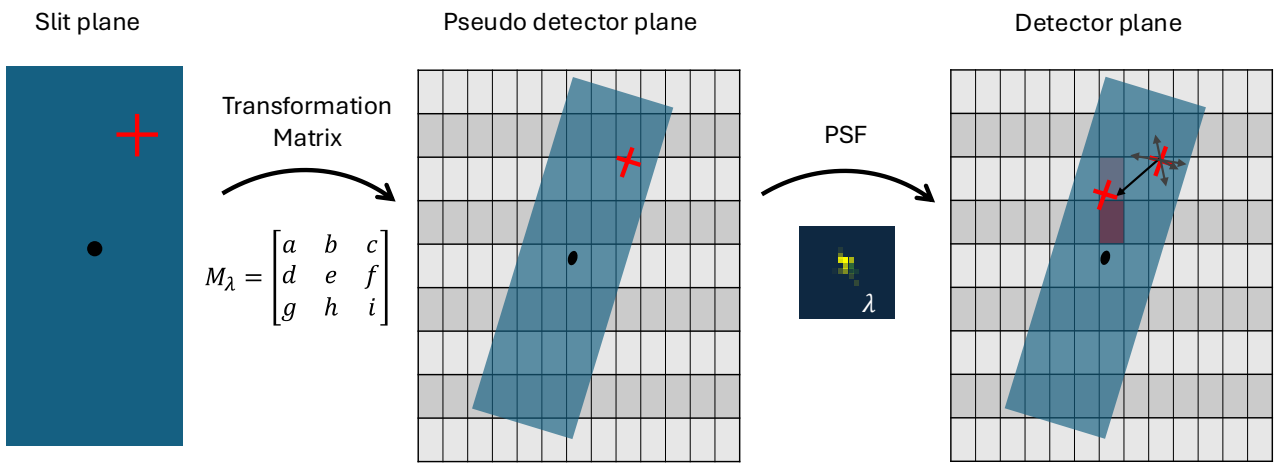

Fig. 9: Pyechelle workflow: the spectrograph optics are modeled as a combination of wavelength-dependent transformation matrices and point spread functions, allowing the simulation of how a given flux at a specific wavelength is mapped onto the spectrograph detector (adapted from Stürmer et al. (2018)).

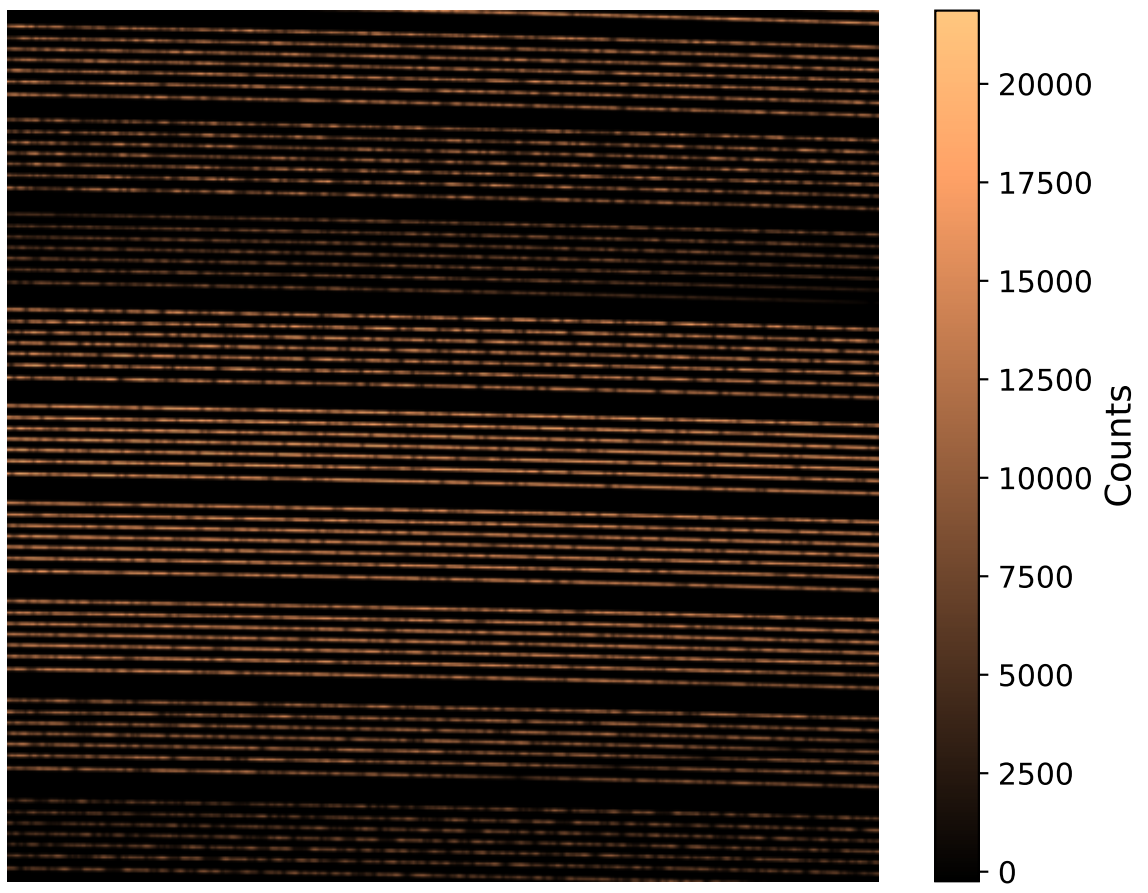

Fig. 10: Simulated 2D raw frame using PyEchelle and the RISTRETTO optical model (zoom on the full frame).

spaxel to another, but it also reduces potential interference between neighboring fibers.

PHOENIX synthetic spectra used for our simulation are expressed in erg/s/cm$^2$/cm at the surface of the star. As input for PyEchelle we need to convert these fluxes to fluxes measured at the telescope entrance and convert units into photons/s/cm. Let $F_{\rm Phoenix}(\lambda)$ denote the Phoenix flux of Proxima. To convert $F_{\rm Phoenix}(\lambda)$ to $F_{\rm Pyechelle}(\lambda)$ in photons/s/cm units at the entrance of the telescope, we performed the calculation:

$$F_{\rm Pyechelle}(\lambda) = F_{\rm Phoenix}(\lambda) \cdot \frac{R_\star^2}{d^2} \cdot \frac{\lambda}{hc} \cdot A_{\rm telescope}. \quad (2)$$

Here, $h$ is the Planck constant, $c$ is the speed of light, $A_{\rm telescope}$ is the collecting area of the telescope (49.3 m$^2$ for the VLT), and $R_\star$ and $d$ are specified in Table 1. The output flux units on the detector are photons/pixel.

## 4. Data analysis methodology

In this section, we describe the data analysis performed on the simulated spectra to detect the exoplanet signal. The process begins with the extraction of the 1D spectrum from the simulated 2D raw frames. We then outline the various steps that can be applied to identify the planetary signal, including methods for distinguishing it from the host star light. We assume that the longitude of the ascending node has been determined. The goal is to demonstrate how the simulated data can be analyzed effectively to confirm the presence of the exoplanet and isolate its spectral features.

### 4.1. Extraction of 1D spectra

To extract the 1D spectrum from the raw frame we employed the optimal extraction technique described in Horne (1986). The optimal extraction method takes into account the cross-dispersion profile of echelle orders to optimize signal-to-noise ratios. Our data reduction approach includes the following main steps:

1. A master flat-field frame was generated by simulating 10 flat-field exposures with a white-light source.
2. The order trace center for each spectral order was fitted using Gaussian functions along the cross-dispersion direction.
3. The master flat-field frame was employed to compute the normalized order profile used in the optimal extraction.
4. We utilized the pre-existing wavelength solution available in the RISTRETTO model. This choice was made for simplicity, as extracting the wavelength solution from a simulated U-Ne lamp and developing a dedicated pipeline was beyond the scope of this work. RISTRETTO is a stabilized instrument and the wavelength solution can be considered stable at the level of a few m/s over 24 hours. Even assuming a velocity shift of a few m/s over an hour, the corresponding flux variation—based on the typical width of a stellar absorption line—would be on the order of a few parts per thousand. This level of variation is about 2 orders of magnitude smaller than the typical measurement uncertainty and can therefore be considered negligible. Furthermore, if the drift evolves linearly with time, its impact is effectively mitigated by our method, which takes the average of the short exposures taken before and after the long exposure.

For the flux error estimation, we employed two distinct methods. For the short-exposure spectra, where the star is positioned in the off-axis spaxel and exhibits a high S/N, we used the errors obtained with the optimal extraction method described in Horne (1986). For long-exposure spectra with lower S/N, a different strategy was employed to have an accurate error estimation. At lower S/Ns, an accurate determination of photon noise becomes difficult. Since the stellar halo dominates the spectra containing the planetary signal, the associated photon noise is primarily linked to the stellar halo flux. Thus, both exposures (star-only and star halo + planet) share the same dominant signal, which differs only in terms of S/N. The flux from the high-S/N exposure was used for estimating the photon noise on the low-S/N spectra, after rescaling using a first-degree polynomial in each order. A similar approach is described in more detail in Bourrier et al. (2024).

All the analyses presented in the next sections were performed using spectra extracted as described above. We chose not to merge the different orders in order to avoid any additional interpolation on a common wavelength grid.

### 4.2. Data preparation

We initially tested a cross-correlation function (CCF) - based approach to identify the planetary signal. However, the main challenge lies in accurately removing the stellar halo signal from the off-axis spaxel spectra, where the planetary signal is present. Each spectrum exhibits broadband flux variations due to the fiber coupling functions and atmospheric conditions (as shown in Figure 8). These uncalibrated spectral variations make it extremely difficult to accurately model or remove the stellar halo alone, resulting in residual stellar contamination that significantly affects the CCF analysis. Therefore, we opted for the direct fitting of a halo+planet model across all spectra simultaneously, which has become the preferred method in the literature (Ruffio et al. 2019; Landman et al. 2024).

This approach first requires a comprehensive description of the off-axis spaxel observations, which include both the stellar halo spectrum and the planetary reflected spectrum, shifted in radial velocity due to the orbital motion of the planet. For each long exposure and each individual spectral order we can write:

$$F_1(\lambda) = T_{1,\mathrm{grey}} \cdot T_{1,\mathrm{tell}}(\lambda) \cdot [\underbrace{\rho_i^s(\boldsymbol{r}_{\mathrm{on}}) \cdot F_s(\lambda)}_{\text{stellar halo}} + \underbrace{\rho_i^p(\boldsymbol{r}) \cdot F_s(\lambda, v_r) \cdot A_g(\lambda, v_r) \cdot g(\alpha) \cdot \frac{R_p^2}{a^2}}_{\text{planet}}]. \quad (3)$$

Here, $F_p(\alpha, \lambda, v_r) = F_s(\lambda, v_r) \cdot A_g(\lambda, v_r) \cdot g(\alpha) \cdot \frac{R_p^2}{a^2}$ from Eq. 1. $T_{1,\mathrm{grey}}$ represents the atmosphere's grey transmission function as well as the instrument transmission, which may both be strongly time-variable but vary slowly with wavelength; we assume here that $T_{\mathrm{grey}}$ is constant over a spectral order. $T_{1,\mathrm{tell}}(\lambda)$ represents telluric molecular absorption. $\rho_i^s(\boldsymbol{r}_{\mathrm{on}})$ and $\rho_i^p(\boldsymbol{r})$ respectively denote the coronagraphic IFU coupling functions of the stellar halo spectrum and the planet spectrum. The subscript indicates the fiber where the coupling is measured. The couplings depend on the position of the source $\boldsymbol{r}$ (star or planet), with $\boldsymbol{r}_{\mathrm{on}}$ meaning that the source is placed on-axis. A deeper discussion about the coefficients $\rho$ is presented in Sect.4.3. $F_s(\lambda)$ denotes the star spectrum, and $v_r$ represents the orbital radial velocity of the planet. $A_g(\lambda)$ represents the albedo spectrum of the planet, $g(\alpha)$ stands for the phase function, $R_p$ denotes the planetary radius, and $a$ indicates the semi-major axis of the planet orbit.

Similarly, we can express the flux in the off-axis spaxel during the short exposures as

$$F_2(\lambda) = T_{2,\mathrm{grey}} \cdot T_{2,\mathrm{tell}}(\lambda) \cdot \rho_i^s(\boldsymbol{r}_{\mathrm{off}}) \cdot F_s(\lambda), \quad (4)$$

where $\boldsymbol{r}_{\mathrm{off}}$ denotes the center of the off-axis spaxel $i$. To model the stellar halo in the long exposures, we decided to take the average of the high-S/N short exposures taken just before and after the long exposure. We assume that bracketing each long exposure in this way provides a reasonable approximation to the barycentric correction and telluric absorption during the long exposure, i.e. $T_{1,\mathrm{tell}}(\lambda) = T_{2,\mathrm{tell}}(\lambda)$. In the following, $F_2(\lambda)$ represents the average of the two bracketing short exposures. We can then compute the ratio between the long and short exposures to calibrate out telluric absorption as

$$\frac{F_1(\lambda)}{F_2(\lambda)} = \frac{T_{1,\mathrm{grey}}}{T_{2,\mathrm{grey}}} \cdot \frac{1}{\rho_i^s(\boldsymbol{r}_{\mathrm{off}})} \cdot [\rho_i^s(\boldsymbol{r}_{\mathrm{on}}) + \rho_i^p(\boldsymbol{r}) \cdot \frac{F_s(\lambda, v_r)}{F_s(\lambda)} \cdot A_g(\lambda, v_r) \cdot g(\alpha) \cdot \frac{R_p^2}{a^2}]. \quad (5)$$

We can next compute the weighted average of this ratio over each spectral order ($o$), where the weight is given by the inverse errors squared:

$$\left\langle \frac{F_1(\lambda)}{F_2(\lambda)} \right\rangle_o = \frac{T_{1,\mathrm{grey}}}{T_{2,\mathrm{grey}}} \left\langle \frac{\rho_i^s(\boldsymbol{r}_\mathrm{on})}{\rho_i^s(\boldsymbol{r}_\mathrm{off})} \cdot \left[ 1 + \frac{\rho_i^p(\boldsymbol{r})}{\rho_i^s(\boldsymbol{r}_\mathrm{on})} \cdot \frac{F_s(\lambda, v_r)}{F_s(\lambda)} \cdot A_g(\lambda, v_r) \cdot g(\alpha) \cdot \frac{R_p^2}{a^2} \right] \right\rangle_o . \quad (6)$$

We can rewrite the first coupling ratio as

$$\frac{\rho_i^s(\boldsymbol{r}_\mathrm{on})}{\rho_i^s(\boldsymbol{r}_\mathrm{off})} = \frac{\rho_i^s(\boldsymbol{r}_\mathrm{on})}{\rho_0^s(\boldsymbol{r}_\mathrm{on})} \cdot \frac{\rho_0^s(\boldsymbol{r}_\mathrm{on})}{\rho_i^s(\boldsymbol{r}_\mathrm{off})}, \quad (7)$$

where $\rho_0^s(\boldsymbol{r}_\mathrm{on})$ denotes the coupling of an on-axis source into the central spaxel. The purpose of this transformation is explained in Sect.4.3.

Finally, we can compute the normalized ratio to eliminate the broadband variations in atmospheric and instrumental transmission:

$$\frac{\frac{F_1(\lambda)}{F_2(\lambda)}}{\left\langle \frac{F_1(\lambda)}{F_2(\lambda)} \right\rangle_o} = \frac{\frac{\rho_i^s(\boldsymbol{r}_\mathrm{on})}{\rho_0^s(\boldsymbol{r}_\mathrm{on})} \cdot \frac{\rho_0^s(\boldsymbol{r}_\mathrm{on})}{\rho_i^s(\boldsymbol{r}_\mathrm{off})} + \frac{\rho_i^p(\boldsymbol{r})}{\rho_i^s(\boldsymbol{r}_\mathrm{off})} \cdot \frac{F_s(\lambda,v_r)}{F_s(\lambda)} \cdot A_g(\lambda, v_r) \cdot g(\alpha) \cdot \frac{R_p^2}{a^2}}{\left\langle \frac{\rho_i^s(\boldsymbol{r}_\mathrm{on})}{\rho_0^s(\boldsymbol{r}_\mathrm{on})} \cdot \frac{\rho_0^s(\boldsymbol{r}_\mathrm{on})}{\rho_i^s(\boldsymbol{r}_\mathrm{off})} + \frac{\rho_i^p(\boldsymbol{r})}{\rho_i^s(\boldsymbol{r}_\mathrm{off})} \cdot \frac{F_s(\lambda,v_r)}{F_s(\lambda)} \cdot A_g(\lambda, v_r) \cdot g(\alpha) \cdot \frac{R_p^2}{a^2} \right\rangle_o} . \quad (8)$$

With this procedure, we eliminate all atmospheric and instrumental effects that cannot be easily modeled or calibrated and do not contain any useful information for the planet detection. This normalized spectrum ratio can be computed for each spectral order in each long exposure, together with its error bars. We will use it as our observable to be modeled in the remainder of this paper.

### *4.3. Data modeling*

We proceeded to model the normalized spectrum ratio using Eq.8. To do so, it's necessary to have suitable models or measurements on hand for the coupling function ratios of Eq. 8, the stellar spectrum $F_s(\lambda)$, and the planet spectrum. We discuss them individually below.

#### 4.3.1. Coupling function ratios

All the coupling function ratios vary as a function of atmospheric conditions (e.g. seeing and elevation) and instrumental parameters. We discuss them individually here below.

The term $\frac{\rho_i^s(\boldsymbol{r}_\mathrm{on})}{\rho_0^s(\boldsymbol{r}_\mathrm{on})}$ represents the coupling function ratio between the off-axis and central spaxels when the star is positioned in the central spaxel. Its value depends on the performance of the AO+coronagraphic system, and thus on seeing conditions and star elevation. We can estimate it directly from the data by computing the median of the ratio between the off-axis and central spaxel spectra in each long exposure in each order. We should point out that the numerator is impacted by the neutral density filter used in the central spaxel to attenuate the stellar PSF, but this can be precisely measured during daytime calibrations. We correct the measured ratio from this factor.

The term $\frac{\rho_0^s(\boldsymbol{r}_\mathrm{on})}{\rho_i^s(\boldsymbol{r}_\mathrm{off})}$ represents the ratio of the coupling efficiencies into the central and off-axis spaxels for sources located at the center of these spaxels, under the same observing conditions. The denominator here comes from the short exposure, but we assume that it differs from the long-exposure value only by a constant factor within a spectral order. Since $\rho_i^s(\boldsymbol{r}_\mathrm{off})$ multiplies all terms in Eq.8, we can absorb this constant factor into the overall normalization. Therefore, for simplicity, we define the denominator as the off-axis spaxel coupling efficiency during the long exposure, as if the star was placed at the center of this off-axis spaxel. This coupling ratio varies by only a few percent with changes in star elevation and seeing, suggesting that its value can be calibrated in the laboratory and remains largely stable in time. In our case, we used a linear fit on the median value in each order.

The term $\frac{\rho_i^p(\boldsymbol{r})}{\rho_i^s(\boldsymbol{r}_\mathrm{off})}$ represents the ratio in coupling efficiency between the true planet position within the off-axis spaxel and the center of that same spaxel. This ratio is highly dependent on the planet orbital parameters, IFU orientation, and epoch of observations. In the context of this paper we simply consider that this term can be constrained to better than ∼10% thanks to the precise knowledge of Proxima b's orbit, including the longitude of the ascending node (see Sect.6). Since its value cannot be estimated directly from the data or from calibrations (contrary to the previous two terms), we do not include this term in the model at this stage (i.e., its value is assumed to be 1).

It is important to note that the accuracy with which we can estimate these ratios directly impacts the precision on the planetary albedo determination. Our science goal is to constrain the albedo with an accuracy of 20%, which implies a combined accuracy from each coupling term in Eq.8 better than this value. Laboratory tests will be carried out with the spectrograph and IFU to demonstrate that this is indeed achievable.

#### 4.3.2. Telluric-free stellar model

To derive the stellar spectrum model $F_s(\lambda)$ directly from the observations, we constructed a high-S/N stellar template free of telluric absorption. This was achieved by selecting, in each spectral bin, the 90th percentile of the flux distribution across all short-exposure spectra, acquired over several months. This approach relies on the large variations in the barycentric correction over an observing season, which leaves most spectral bins in the stellar rest frame free from telluric absorption in at least some of the observations. An example of the obtained stellar template, along with its effectiveness in removing telluric lines, is shown in Fig.11. The template is constructed in the stellar rest frame and Doppler-shifted to the observer's rest frame for each long exposure, considering the appropriate barycentric correction and stellar orbital radial velocity.

#### 4.3.3. Planet model

We introduce below five planet models that will be used to fit the normalized spectrum ratio in Eq.8, in order of increasing complexity:

No planet: The simplest model assumes that the planet is not present or does not reflect any light (null albedo). In this case, Eq.8 becomes:

$$\frac{\frac{F_1(\lambda)}{F_2(\lambda)}}{\left\langle \frac{F_1(\lambda)}{F_2(\lambda)} \right\rangle_o} = \frac{\frac{\rho_i^s(\boldsymbol{r}_\mathrm{on})}{\rho_0^s(\boldsymbol{r}_\mathrm{on})} \cdot \frac{\rho_0^s(\boldsymbol{r}_\mathrm{on})}{\rho_i^s(\boldsymbol{r}_\mathrm{off})}}{\left\langle \frac{\rho_i^s(\boldsymbol{r}_\mathrm{on})}{\rho_0^s(\boldsymbol{r}_\mathrm{on})} \cdot \frac{\rho_0^s(\boldsymbol{r}_\mathrm{on})}{\rho_i^s(\boldsymbol{r}_\mathrm{off})} \right\rangle_o}, \quad (9)$$

where there are no free parameters.

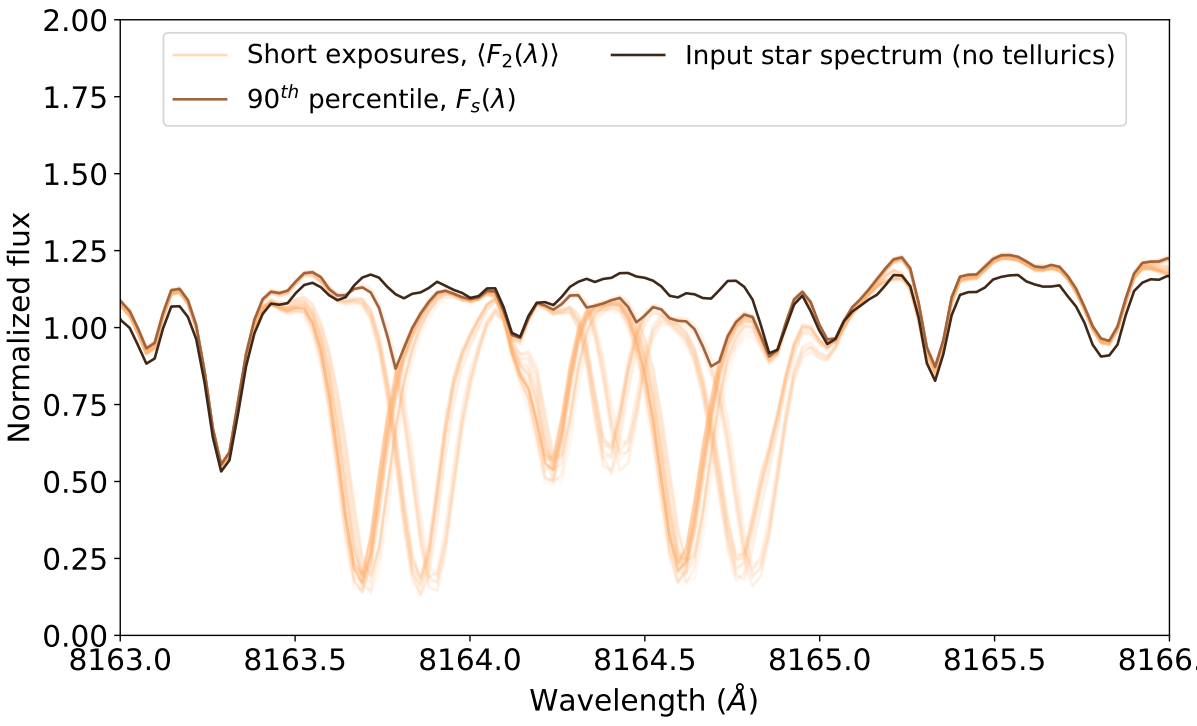

Fig. 11: Creation of the stellar model spectrum (brown) from individual short exposures (orange) with the tellurics affected, and comparison with the true stellar spectrum (black).

Constant albedo: The second simplest model assumes a constant and achromatic planet-to-star flux ratio for the planet, namely that $A_g(\lambda, r_v) \cdot g(\alpha) \cdot \frac{R_p^2}{a^2} = p$, where $p$ is a constant. Therefore, $p$ does not depend on wavelength nor planet orbital phase. Eq.8 then becomes

$$\frac{\frac{F_1(\lambda)}{F_2(\lambda)}}{\left\langle \frac{F_1(\lambda)}{F_2(\lambda)} \right\rangle_o} = \frac{\frac{\rho_i^s(\boldsymbol{r}_{\rm on})}{\rho_0^s(\boldsymbol{r}_{\rm on})} \cdot \frac{\rho_0^s(\boldsymbol{r}_{\rm on})}{\rho_i^s(\boldsymbol{r}_{\rm off})} + \frac{F_s(\lambda, v_r)}{F_s(\lambda)} \cdot p}{\left\langle \frac{\rho_i^s(\boldsymbol{r}_{\rm on})}{\rho_0^s(\boldsymbol{r}_{\rm on})} \cdot \frac{\rho_0^s(\boldsymbol{r}_{\rm on})}{\rho_i^s(\boldsymbol{r}_{\rm off})} + \frac{F_s(\lambda, v_r)}{F_s(\lambda)} \cdot p \right\rangle_o}. \quad (10)$$

In this case, the number of free parameters in the fit is 2: $p$ and the orbital inclination, $i$, which defines all the individual $v_r$ values (one for each long exposure). The other parameters of the Keplerian are already constrained by historical radial velocity measurements and the procedure shown in Sect.6.

Chromatic albedo: A more complex model considers a wavelength-dependent $p$ by allowing it to vary across a finite number of broad wavelength regions. Here $p$ is given as a piecewise function defined over specific wavelength regions:

$$p(\lambda) = \begin{cases} p_1 & \text{if } \lambda \in [\lambda_1, \lambda_2], \\ p_2 & \text{if } \lambda \in [\lambda_2, \lambda_3], \\ \vdots & \vdots \\ p_n & \text{if } \lambda \in [\lambda_n, \lambda_{n+1}]. \end{cases} \quad (11)$$

In this formulation, $p_1, p_2, \ldots, p_n$ are free parameters to be fitted in each wavelength region. The idea behind this more complex model is to fit a planetary albedo that varies with wavelength — a chromatic albedo. This approach avoids making any assumptions about the composition of the planet atmosphere or surface, relying purely on observations. The number of parameters is n + 1: the number of broad wavelength regions and the orbital inclination. For simplicity, as broad wavelength regions, we used equally spaced intervals across the RISTRETTO spectral domain.

Full albedo spectrum: The most complex and comprehensive model assumes a detailed representation of the planetary atmosphere, including a high-resolution albedo spectrum. However, incorporating the full GCM and performing direct Bayesian inference is computationally prohibitive. Therefore, in our case, we only consider the exact same albedo spectrum that was used to generate the simulated spectra (upper plot of Fig.3). A multiplicative term is added as a free parameter to allow for a global scaling of the albedo spectrum. The full albedo model thus has two free parameters (orbital inclination and albedo scaling), similarly to the constant albedo model. Our goal is to explore whether this more sophisticated - and true (in our case) - model is statistically favored over simpler models as a function of the total observing time, potentially providing constraints on the molecular composition of the planetary atmosphere.

Molecular absorption model: A simpler approach to search for molecular signatures is to fit a single-layer, single-molecule absorption model that depends only on temperature, pressure, and column density of absorbing molecules. In this work, we used molecular line data from the HITRAN database (Gordon et al. 2022) via the HAPI Python package (Kochanov et al. 2016) to simulate the absorption features of selected species. The model becomes:

$$\frac{\frac{F_1(\lambda)}{F_2(\lambda)}}{\left\langle \frac{F_1(\lambda)}{F_2(\lambda)} \right\rangle_o} = \frac{\frac{\rho_i^s(\boldsymbol{r}_{\rm on})}{\rho_0^s(\boldsymbol{r}_{\rm on})} \cdot \frac{\rho_0^s(\boldsymbol{r}_{\rm on})}{\rho_i^s(\boldsymbol{r}_{\rm off})} + \frac{F_s(\lambda, v_r)}{F_s(\lambda)} \cdot A_{g,\rm mol}(\lambda, N, T, P, v_r) \cdot p}{\left\langle \frac{\rho_i^s(\boldsymbol{r}_{\rm on})}{\rho_0^s(\boldsymbol{r}_{\rm on})} \cdot \frac{\rho_0^s(\boldsymbol{r}_{\rm on})}{\rho_i^s(\boldsymbol{r}_{\rm off})} + \frac{F_s(\lambda, v_r)}{F_s(\lambda)} \cdot A_{g,\rm mol}(\lambda, N, T, P, v_r) \cdot p \right\rangle_o}, \quad (12)$$

where $A_{g,\rm mol}(\lambda, N, T, P, v_r)$ is the molecular absorption spectrum, which depends on the total column density $N$, the temperature $T$, and the pressure $P$. A homogeneous layer of absorbing gas is assumed here, such that:

$$A_{g,\rm mol}(\lambda, v_r, N, T, P) = \exp\left(-N \cdot \sigma_{\rm mol}(\lambda, v_r, T, P)\right). \quad (13)$$

We modeled the total absorption cross-section $\sigma_{\rm mol}$ using a sum of Voigt profiles over all individual transitions of the considered molecule in the HITRAN database. This adds three degrees of freedom to the model, allowing us to test whether the inclusion of a specific molecule improves the fit compared to the constant albedo model. We are aware that this simple molecular spectrum differs from the input one, which was generated using a full GCM for Proxima b. However, our goal here is not to reproduce the exact input spectrum, but to assess whether a minimal molecular absorption model, reproducing the general high-resolution band structure of a given molecule, is able to improve the fit so as to yield a significant detection of the molecule. An example of model $O_2$ spectra showing how pressure and column density impact the shape of the molecular bands is shown in Fig.12.

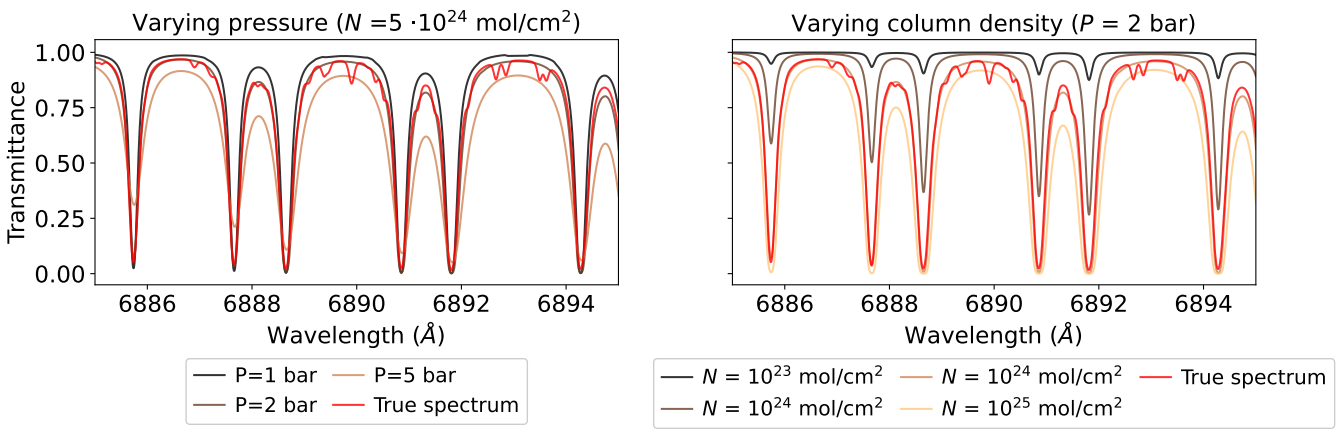

Fig. 12: Single-layer absorption spectra from molecular oxygen for varying pressures and column densities. The full albedo spectrum generated from the GCM modeling of an Earth-like Proxima b is overplotted for comparison. The additional spectral features present in the GCM spectrum but absent in the molecular oxygen spectra are due to $H_2O$ absorption lines.

### 4.4. Model fitting

Bayesian inference has become an essential tool in modern astronomy, enabling researchers to perform both model selection and parameter estimation (Trotta 2008). We adopted this framework here to evaluate which of the five models described in the previous section is statistically favored.

To carry out the Bayesian inference on our synthetic observations, we employed the nested sampling technique (Skilling 2006), which is particularly effective for Bayesian model comparison. Nested sampling is an algorithm designed to compute the Bayesian evidence, denoted by $Z$, while also providing the posterior distribution of the model parameters as a by-product. The Bayesian evidence is the key quantity for robust statistical model comparison (Trotta 2008). In this work, we derive posterior probability distributions and the Bayesian evidence with the nested sampling Monte Carlo algorithm MLFriends (Buchner 2016, 2019) using the UltraNest package (Buchner 2021).

The ratio of model evidence, known as the Bayes factor, allows us to assess the relative support for competing models. Specifically, larger values of the Bayes factor indicate stronger evidence in favor of the model with the higher evidence value. This approach is particularly advantageous in balancing model fit and complexity, as the evidence inherently penalizes overfitting by integrating over the entire parameter space, naturally favoring simpler models unless additional complexity is warranted by the data. According to an empirical scale for evaluating the strength of evidence when comparing two models (Trotta 2008), if the logarithm of the Bayes factor, hereafter denoted $\Delta \ln Z$, exceeds 5, the model with the higher $\ln Z$ is strongly favored over the other. In this work, we evaluate all the models described above to assess which is statistically favored as a function of observing time and conditions. In our case, a Gaussian log-likelihood function is used:

$$\ln L = \sum_{n_0} \sum_{n_p} \sum_{n_e} \left[ -\ln\left(\sigma \sqrt{2\pi}\right) - \frac{(D-M)^2}{2\sigma^2} \right], \tag{14}$$

where $M$ represents our model (right side of Eq.8, with the various approaches discussed in the previous Section 4.3.3) and $D$ our data, i.e. the normalized spectrum ratios $\frac{F_1/F_2}{\langle F_1/F_2 \rangle}$. The total number of data points depends on the number of spectral orders $n_o$, the number of data points per order (4096 pixels = 4096 points) $n_p$, and the number of long exposures $n_e$. Therefore $n_{tot} = n_o \cdot n_p \cdot n_e \sim 10^7$ typically (for 100 exposures).

## 5. Results

### 5.1. Impact of seeing and elevation on data quality

Observing conditions (most notably airmass and seeing) will significantly impact the performance of RISTRETTO. The quality of the AO correction directly depends on the magnitude of the atmospheric turbulence along the line of sight to the target. Poor seeing conditions and low elevations will degrade both the Strehl ratio (thus the coupling efficiency into the IFU spaxels) and the achievable raw contrast at 2 $\lambda/D$. We may therefore have to set upper limits on seeing and airmass beyond which RISTRETTO observations would become inefficient. We performed some tests considering the elevation of Proxima Centauri and seeing probability distributions at Paranal, Chile, where the VLT is located. Fig.13 shows the effective seeing distribution at an elevation of 45° (typical for Proxima), and the histogram of the elevation itself (for elevations > 30°), as seen from Paranal.

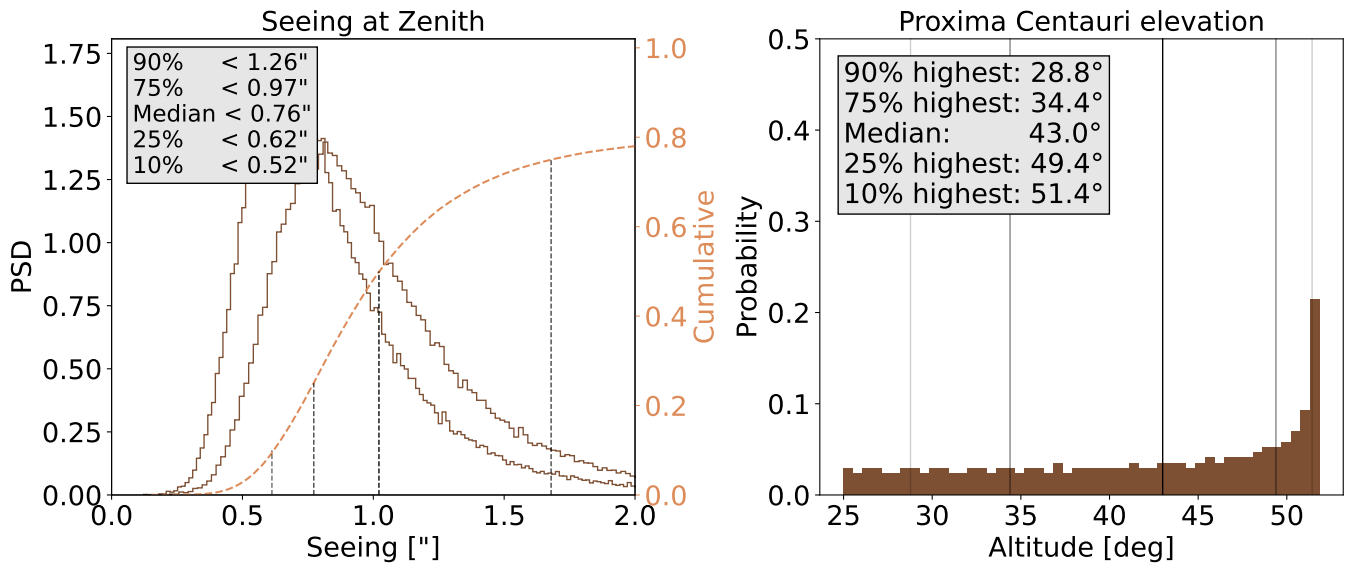

Fig. 13: Effective seeing (at elevation 45°) and elevation distributions for Proxima Centauri at Paranal Observatory, Chile.

We considered possible combinations of five elevations (38°, 43°, 48°, 52°, and 90°, with 90° serving as a reference) and five seeing conditions (0.52, 0.62, 0.76, 0.97, and 1.26 arcseconds, corresponding to the 10th, 25th, 50th, 75th, and 90th percentiles of the Paranal seeing distribution at zenith), resulting in 25 possible combinations. For each of these 25 combinations, we used the appropriate IFU coupling functions as described in Sect.3.5. The mean coupling functions of the stellar halo (transmission in the off-axis spaxels, with the object placed in the central spaxel) are shown in Figure 14.

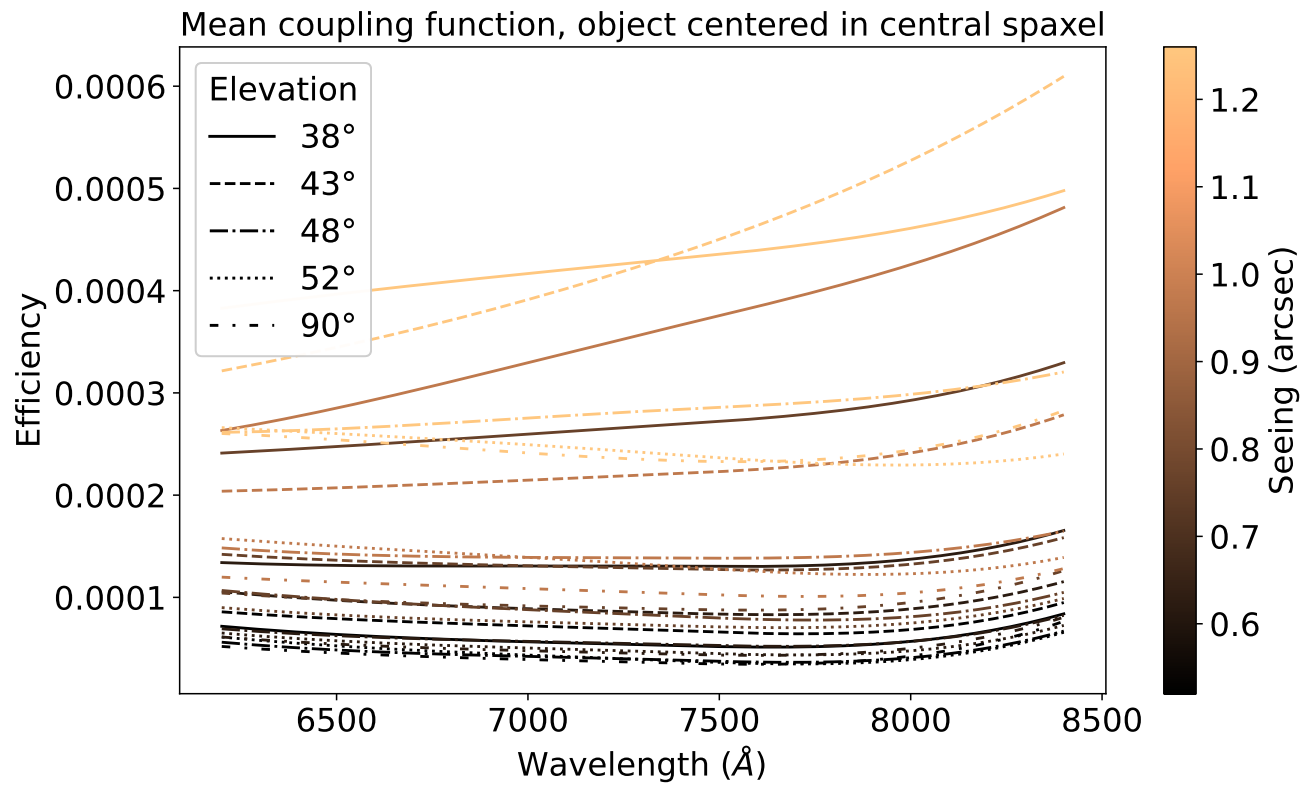

Fig. 14: Average off-axis coupling functions, when an object is placed in the central spaxel, for different elevation and seeing conditions.

We then computed the wavelength-averaged value of the ratio $\rho_i^p(\boldsymbol{r})/\sqrt{\rho_i^s(\boldsymbol{r}_{\rm on})}$, which is proportional to the expected S/N on the planet spectrum in the photon-limited regime (Lovis et al. 2017). We found that observations with a seeing at zenith of 1.26 arcsec yield an expected planet S/N that is 2.5 times lower than in median seeing conditions (0.76 arcsec at zenith). Compensating for this poor S/N would require increasing the exposure time by a factor $\sim$ 6, making such observations inefficient.

We thus decided to restrict observations to seeing conditions better than 0.97 arcsec at zenith (i.e., the 75th percentile). Similarly, since we have a sufficient total observing time available during an observing season, we require an elevation higher than 36°, corresponding to an airmass of 1.7. Moreover, the proposed strategy is to observe for a greater number of hours than strictly necessary, allowing for the selection of the best observations a posteriori. In fact, after each exposure, we can evaluate the performance of the AO + coronagraph system. Consequently, our strategy is to observe Proxima b only when the conditions described above are met, and then select a subset of the best ex-posures (e.g., the best 90% or 95%) for subsequent analysis and planet detection.

### 5.2. Planet model fits and model comparison

Below, we present the nested sampling fits of the different models described in Section 4.3.3 and perform a Bayesian model comparison. This analysis allows us to evaluate which model is statistically preferred and provides insights into the detectability of the planet itself and the molecular species in the planetary atmosphere.

#### 5.2.1. Detectability of Proxima b: no planet versus constant albedo model

First, we simulated multiple sets of observations and seeing distributions to determine the minimum number of telescope hours required to detect an Earth-like Proxima b at high significance. Formally, planet detection can be assessed through a Bayesian model comparison between planet models 1 (null albedo) and 2 (constant albedo) described in Sect.4.3.3. In practice, we found that requiring $\Delta \ln Z > 5$ in favor of the constant albedo model is generally not sufficient to obtain well-behaved posterior distributions for the two free parameters of the model (orbital inclination and planet-to-star flux ratio). Instead, we established a detection criterion based on the achieved precision on the flux ratio, which is more stringent than simple planet detection, but scientifically more informative. A 20% precision on the albedo is sufficient to place meaningful constraints on the planetary atmosphere and we used this value as a threshold.

With this criterion and after performing ten sets of simulations, we found that the minimum number of hours (= long exposures) to detect the planet is 55 hours, of which the best 50 hours are used. We note that with such an exposure time the Bayes factor $\Delta \ln Z$ in favor of the constant albedo model versus the no-planet model is 10.5 on average, meaning that the former model is favored over the latter by a factor of about 36,500 to 1. Furthermore, using the best 80 hours out of 75, it is possible to reach a precision of 15% on the flux ratio.

It is interesting to note that 55 hours of total exposure time corresponds to a cumulative signal-to-noise ratio of $\sim$ 150 per spectral bin, with the stellar halo as the dominant signal. At the level of each exposure, under median seeing and elevation conditions, the stellar photon noise per spectral bin is $\sim$ 20 electrons/hour, while the read-out noise is $\sim$ 5.5 electrons. We are therefore still in the photon-limited regime despite the low signal level.

Fig.15 illustrates the evolution of $\Delta \ln Z$ as exposure time increases. The solid line represents the mean value across the 10 simulated datasets, while the shaded region indicates the corresponding scatter. The scatter in $\Delta \ln Z$ between the sets is approximately 5.6 for 55 hours of observations. Variations in the seeing distribution are not the cause of the scatter between datasets, as the mean IFU coupling functions, which depend on seeing and elevation distributions, only vary by a few percent across the sets, thanks to the best exposures choice. As a test, we ran multiple simulations using the same seeing distribution but different noise realizations, and observed a similar level of scatter. This suggests that the noise realization is the main cause of the scatter in $\Delta \ln Z$ between simulation sets.

An example of the nested sampling posterior distributions for the constant albedo model (Eq.10) is shown in Fig.16 for 55 hours of observations. The results for all 10 simulations are shown in Fig.17. As can be seen from the figure, the mean values and widths of the posterior distributions match the expected "true" (i.e. simulated) values and uncertainties for both parameters. Moreover, the standard deviation of the median flux-ratio values across the ten posterior results is comparable to the standard deviation within a single posterior distribution, indicating the robustness of our results. The mean flux ratio is underestimated by ~20% because we have not accounted for the third coupling function ratio in our modeling, as explained in Sect.4.3.1. In the simulations, this term has a typical value of ~0.86. Setting it to 1.0 in the modeling induces a 20% underestimation of the flux ratio, which we indeed find in our results.

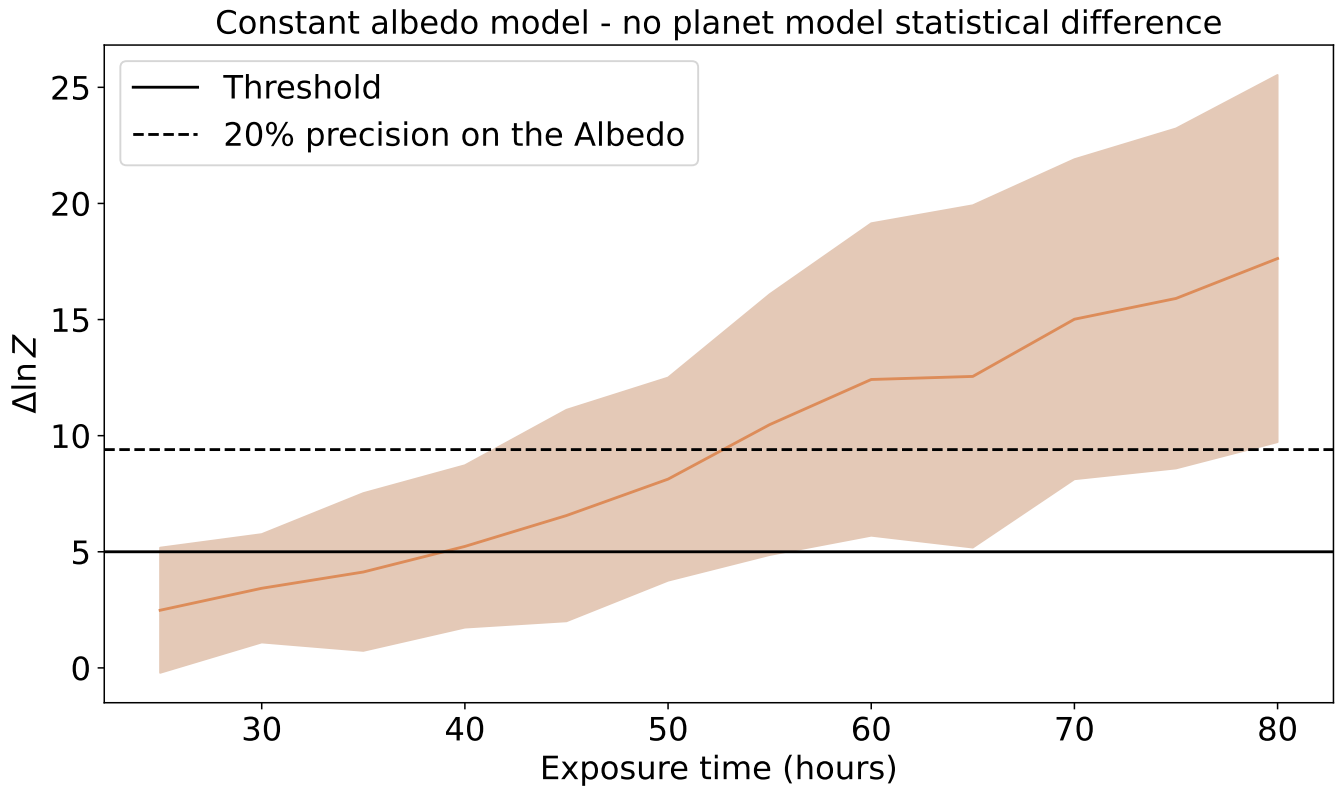

Fig. 15: $\Delta \ln Z$ between the model with constant albedo and the model without the planet as a function of exposure time. The solid line represents the mean across the 10 simulated datasets, while the shaded region indicates the corresponding scatter.

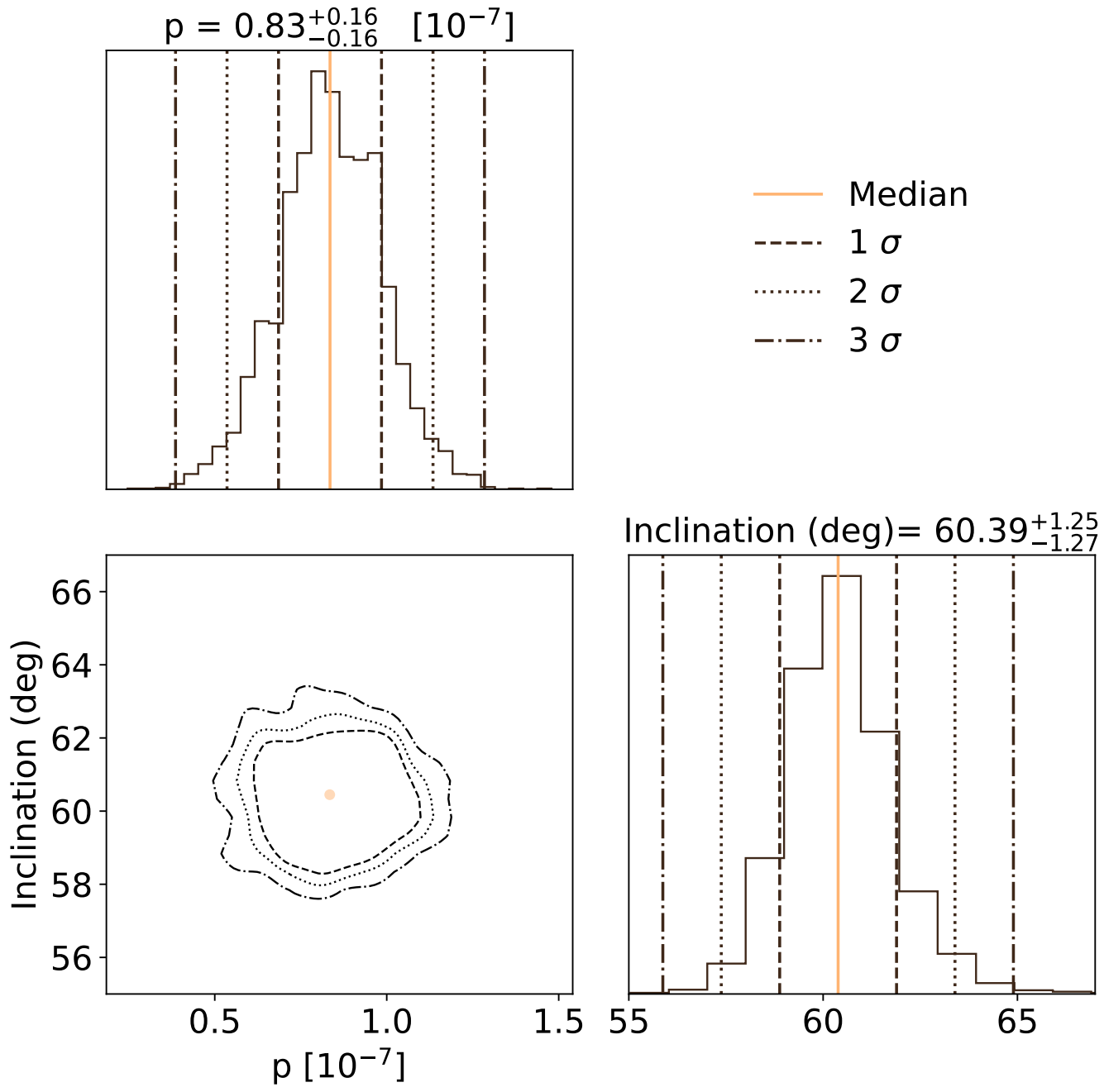

Fig. 16: Example of the posterior distributions for the orbital inclination and planet-to-star flux ratio obtained with the constant albedo model. A relative precision of ~20% is achieved on the flux ratio in 55 hours of exposure time.

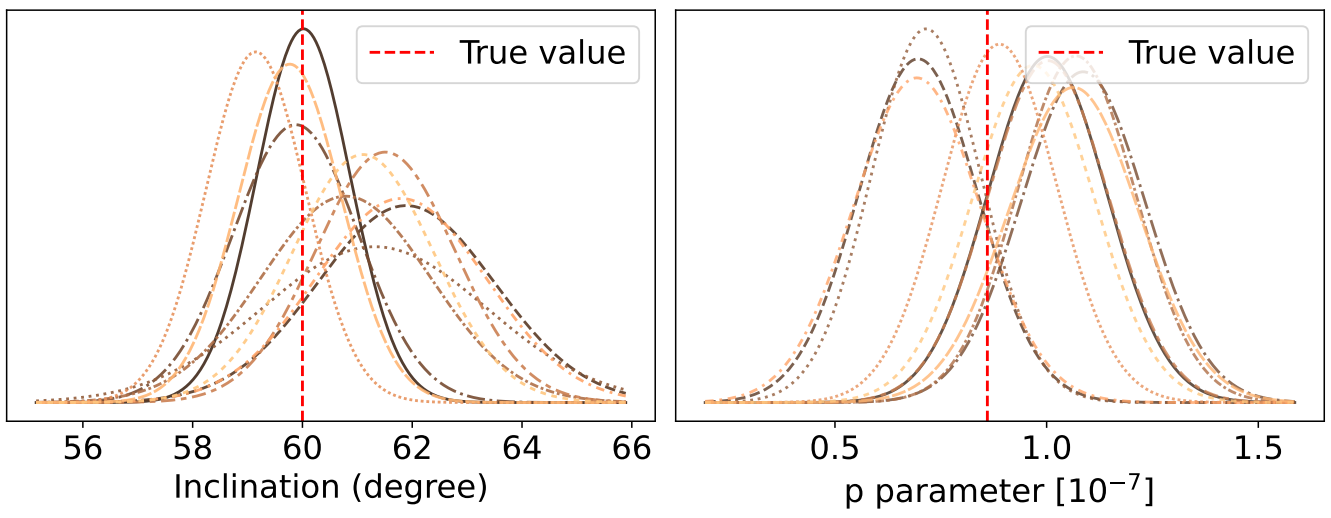

Fig. 17: Posterior distributions (approximated by Gaussian functions) for the constant albedo model for the 10 sets of simulations. Left: Orbital inclination. Right: Planet-to-star flux ratio.

#### 5.2.2. Constant albedo versus chromatic albedo models

Using 55 hours of observing time (best 50 exposures), we were able to compare the chromatic albedo model from Sect.4.3.3 (Eq.11) with the constant albedo model. We considered a number of $p$ parameters ranging from 2 to 10, using equal-sized broadband wavelength bins over the RISTRETTO wavelength range. For computational reasons, we performed the nested sampling fit over the 10 simulation sets for all models but the ten-parameter one, which we fit to only two sets. The mean $\Delta \ln Z$ over all the simulated sets is shown in Table 2.

Table 2: $\Delta \ln Z$ for models with different number of parameters

| # of parameters $p_n$ | 0 | 1 | 2 | 3 | 5 | 10 |
|---|---|---|---|---|---|---|
| $\ln Z_n - \ln Z_1$ | -10.5 | 0 | -1.43 | -1.89 | 2.6 | -1.48 |

**Notes.** $n = 0$ corresponds to no planet, $n = 1$ to a constant albedo, and $n \geq 2$ to models with a chromatic albedo.

Table2 shows that chromatic albedo models are not strongly statistically favored over the constant albedo model. Among the chromatic albedo models, the standard deviation of $\Delta \ln Z$ across simulated datasets is approximately 2–3. To illustrate the results, Fig.18 shows the fitted 5-parameter model overlaid on the "true" albedo spectrum, using an exposure time of 55 hours. From Fig.18 it is clear that the mean of the actual albedo spectrum (squared markers) is very close to the continuum value. All fitted parameters are within 2-3 $\sigma$ of the true values. This method is effectively equivalent to a low-resolution fit; however, since the mean values are in this case close to the continuum, the chromatic model is not statistically preferred. If we considered a different planetary composition, such as one with higher atmospheric pressure, the absorption bands could become saturated. This would lower the mean spectral flux in those bands, potentially making the chromatic albedo model statistically favored.

#### 5.2.3. Constant albedo versus full albedo spectrum

In the specific case of an Earth-like Proxima b, the dominant contribution to the reflected planetary spectrum comes from the stellar absorption lines reflected off the planetary atmosphere, which span the entire wavelength range. In contrast, molecular absorption features (e.g., those from $H_2O$ and $O_2$) are confined to only a few spectral orders, making them inherently more difficult to detect. To probe molecular absorption from the planetary atmosphere, we compared the constant albedo model with the full albedo model, which incorporates the high-resolution albedo spectrum used to generate the synthetic data. The resulting $\Delta \ln Z$ between the two models is shown in Fig.19 for the ten simulated datasets.

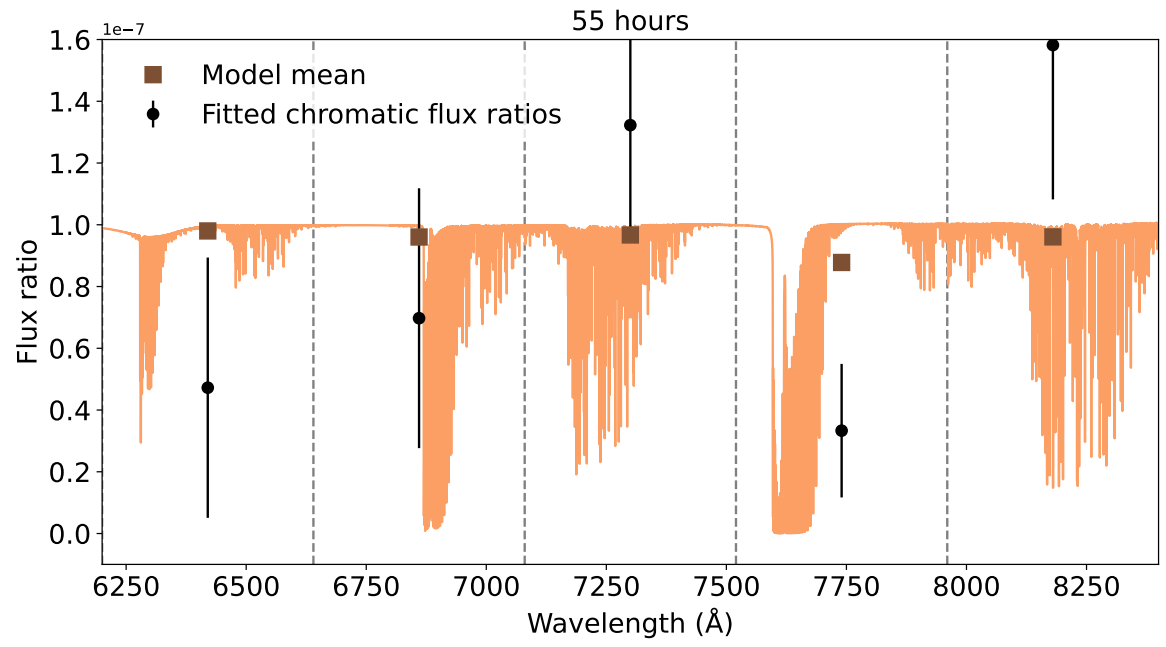

Fig. 18: Fitted five-parameter chromatic albedo model using 55 hours of exposure time.

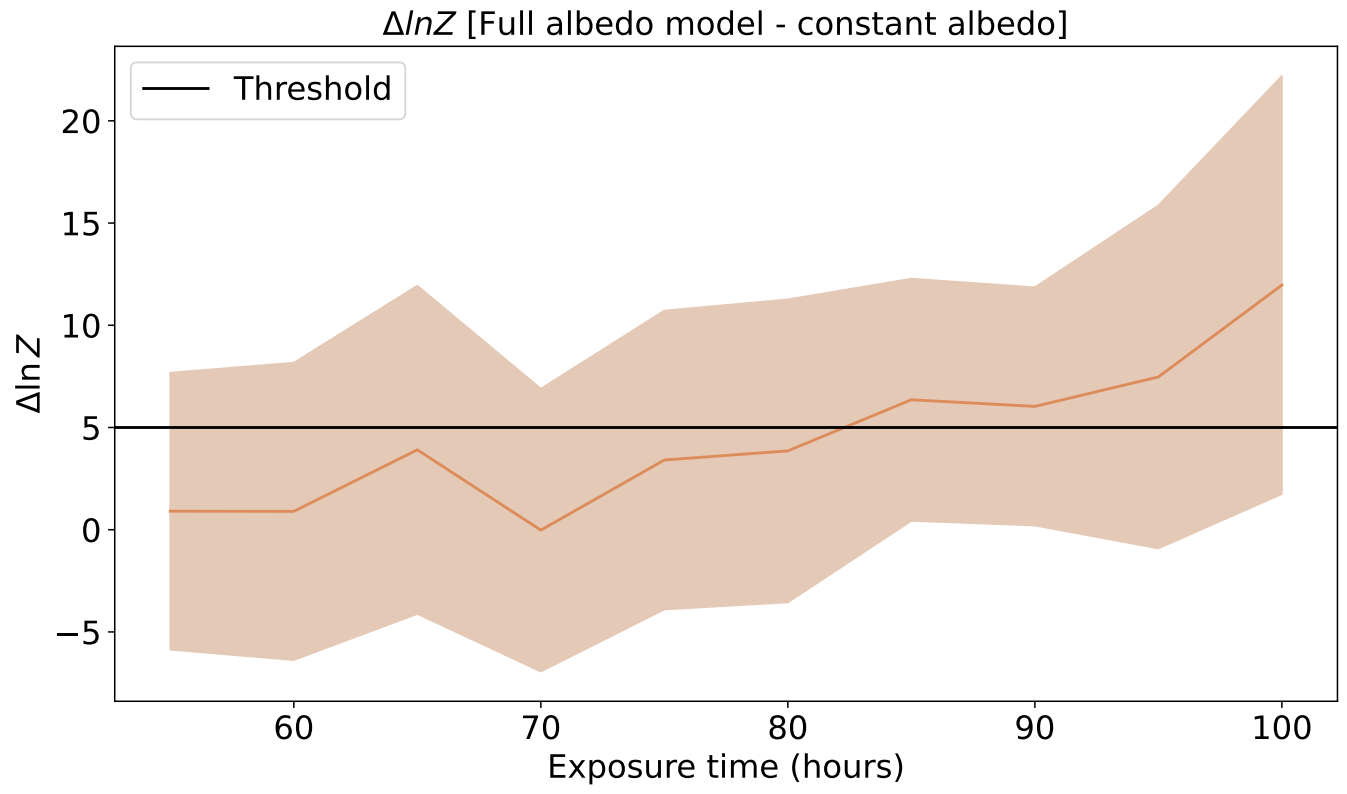

Fig. 19: $\Delta \ln Z$ between the full albedo model and the constant albedo model as a function of exposure time. The solid line represents the mean across the ten simulated datasets, while the shaded region indicates the corresponding scatter.

In Fig.19, we show that the molecular signatures in the planetary spectrum become distinguishable ($\Delta \ln Z > 5$) after approximately 80–85 hours of exposure time. This confirms that the largest part of the planetary signal comes from the stellar lines reflected by the planet, which can be detected after just 55 hours. To reveal the molecular absorption features of an Earth-like atmosphere, at least 30 more hours of exposure are required.

We must point out that the full albedo model includes only two free parameters, which does not reflect its true complexity. Important factors such as atmospheric pressure, temperature profile, molecular abundances, cloud properties, and surface characteristics are not explored. The actual parameter space of the model has been reduced to two dimensions because we have access to the "true" planetary albedo spectrum (i.e., the one we used to build our synthetic observations). A full exploration of the GCM parameter space—through retrieval techniques or grid sampling—is well beyond the scope of this work. Nonetheless, the results suggest that it should be possible to extract significant information on molecular absorption in about 100 hours of RISTRETTO observations for an Earth-like Proxima b.

#### 5.2.4. Constant albedo versus molecular absorption model

Finally, we fitted the data with the simple molecular absorption model described in Sect.4.3.3. We performed nested sam-pling runs with 100 hours of exposure time for the 10 simulated datasets. Free parameters were: orbital inclination, albedo scaling factor, total column density, and pressure. The temperature was kept fixed at 250 K for simplicity, after noting that it has only a marginal impact on the fit when varied within physically plausible limits.

We fit $O_2$ and $H_2O$ separately. For $O_2$, the resulting $\Delta \ln Z$ are 22 on average in favor of the molecular model, thus yielding a significant detection of $O_2$ in 100 hours. However, the resulting posterior distributions for the column density are generally broad and poorly constrained. These results hint at the fact that our molecular model is too simplistic to provide physically meaningful results, which is not unexpected given the much more complex atmospheric structure that is produced by the GCM. It is nevertheless useful as a first-order tool to probe the existence of a given molecule in the atmosphere. A representative set of posterior distributions for the $O_2$ model is shown in Appendix C. For the $H_2O$ model, the results yield a $\Delta \ln Z$ close to zero when compared to the constant albedo model, indicating no significant preference for the presence of water vapor after 100 hours of exposure. This suggests that a longer observation time may be necessary to detect $H_2O$, whose absorption lines are shallower and more difficult to distinguish. As with $O_2$, the posterior distributions for column density and pressure remain poorly constrained and should not be interpreted as physically meaningful. A set of posterior distributions for the $H_2O$ model is shown in Appendix C.

### 5.3. Albedo determination

All planet models considered above have as free parameters the planet-to-star flux ratio(s), not the planetary albedo(s) directly. Isolating the planetary atmosphere and surface contributions, through the albedo and phase function, require knowledge of the planetary radius and orbital semi-major axis. Moreover, the model in Eq.8 contains the coupling function ratios. These must also be known accurately to avoid biasing the flux ratio measurements. We thus examine here how accurately these factors can be determined.

Reflected-light observations exhibit an intrinsic degeneracy between albedo and planetary radius. For transiting planets the radius can be directly measured, while for hot planets observed in thermal emission the radius may be determined through the Stefan-Boltzmann law. Unfortunately, no such independent radius measurement is available for cool, non-transiting planets such as Proxima b. However, we may estimate the radius of individual exoplanets in an indirect way, using empirical mass-radius relations derived from statistical population studies. Once RISTRETTO has measured the orbital inclination of Proxima b, its true mass will be known and can be used to infer its likely radius. The planet is very likely to fall in a mass regime where most if not all known planets are rocky in composition. Empirical mass-radius relations in this regime are quite tight, with an observed spread in radius at the 3-5% level at a given mass (Parc et al. 2024). We may thus expect to obtain a radius-squared estimate for Proxima b (the relevant quantity for reflected light) that has an accuracy of 6-10%. While this is not ideal, it appears still good enough to usefully constrain the albedo.

The coupling function ratios are discussed in Sect.4.3.1. They can be decomposed into three terms, the first two of which can be directly measured with an accuracy of a few percent. The third term is more problematic, as it requires a precise knowledge of the planet position within the IFU at each observing epoch. The four orbital elements that can be derived from RV measurements are now known to sufficient accuracy to not be a limiting factor in the case of Proxima b (Mascareño et al. 2025). The orbital inclination will be accurately measured by RISTRETTO. Thus the only remaining source of significant uncertainty is likely to be the longitude of the ascending node. Below, we present an original method to constrain this parameter directly from RISTRETTO observations.

## 6. Constraining the longitude of the ascending node using differential limb coupling

The single-mode fibers used in RISTRETTO induce strong coupling variations across the IFU field of view (see Figs.8). In particular, obtaining a homogeneous coverage of the annulus at 2 $\lambda/D$ requires two exposures with a 30 deg rotation of the IFU, reducing the effective transmission by a factor of 2. It would thus be advantageous to know a priori the longitude of the ascending node of the planetary orbit, so that the IFU can be optimally oriented. Modern planet formation theories suggest that, for systems hosting several close-in small planets, the stellar rotational axis often aligns with the orbital plane of the planets (Winn & Fabrycky 2015), namely, the obliquity of the star is small. Thus, measuring the orientation of the stellar rotation axis could provide an important constraint on the longitude of the ascending node. In this section, we outline a method for measuring the star rotational axis, which we then assume to coincide with the longitude of the ascending node of Proxima b's orbit. This method is based on a concept outlined in Guyon et al. (2024) and Baker et al. (2024).

Stars are not point sources; they have a physical extent and undergo rotation, which causes varying radial velocities across their surfaces. This rotation effect can be exploited using the steep coupling gradient in a single-mode fiber to measure the star rotational axis. Specifically, the coupling into the central fiber of RISTRETTO is illustrated in Figure 20 together with a rotating stellar disk. By placing a star of a finite angular dimension where the coupling gradient is the steepest, we can induce an offset in the measured radial velocity of the star. This offset results from the difference in radial velocities between opposite sides of the stellar disk, combined with the differential coupling efficiency of the fiber. For instance, in Figure 20, we will measure a redshift in the stellar spectrum because the redshifted hemisphere is closer to the fiber center and thus has a higher coupling efficiency.

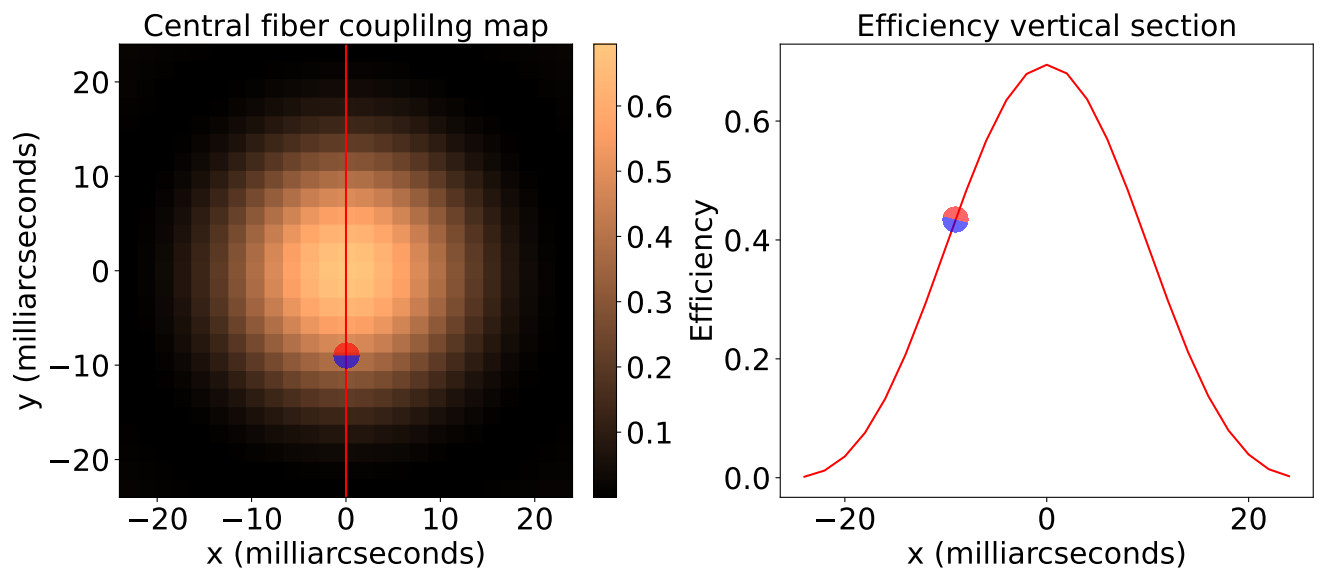

Fig. 20: Coupling map of the central RISTRETTO fiber and a cross-section through it. The red-blue circle schematically represents the rotating, Doppler-shifted surface of the star. The stellar disk size approximately matches the angular diameter of Proxima (1 mas).

By positioning the star at a fixed distance from the center of the fiber and rotating it around that center, the variation in radial velocities across the stellar disk can be leveraged to determine the orientation of its spin axis. As an example, in the right plot of Figure 21, if we place the star at positions A, B, C, and D, and then compute the radial velocity difference between the spectra taken at A-B and C-D, we will observe a difference between C and D but not between A and B. This suggests that the spin axis is likely close to the A-B direction.

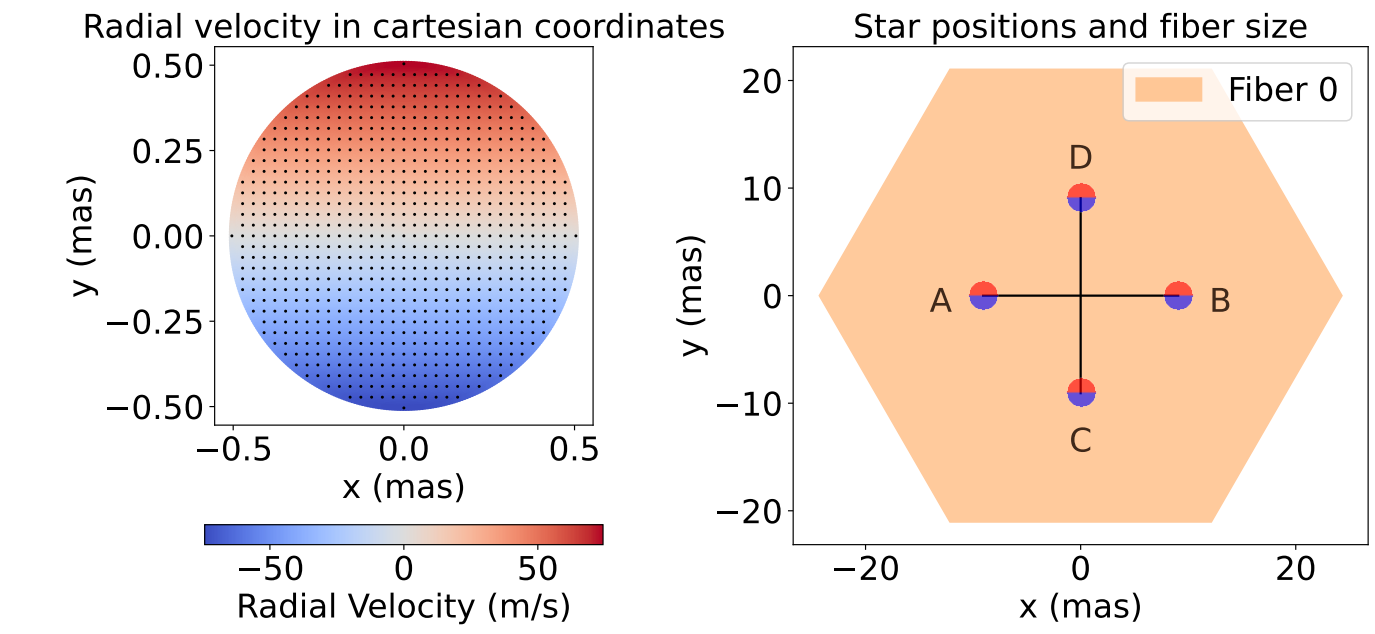

Fig. 21: Left panel: Radial velocity map on the surface of Proxima Centauri and selected grid points to compute the disk-integrated spectra. Right panel: Example of pairs of opposite exposures.

In our simulations, we first identified the optimal distance from the fiber center to achieve the steepest coupling gradient. From the coupling map in Figure 20, the distance yielding the steepest gradient was determined to be $r = 9$ mas. We constructed a radial velocity map for the stellar surface based on the known stellar radius, rotation period, and distance to the system. The map is shown in the left plot of Figure 21. The star was then positioned at 9 mas from the fiber center, and its position angle $\theta$ was varied between zero and $2\pi$ in 32 steps (in Figure 21, A is at $\theta = 180°$, B at $\theta = 0°$ etc.). At each position angle, we computed a weighted average of the stellar spectrum using the radial velocity and coupling maps over a grid of about 800 points on the stellar surface (black points in the left panel of Figure 21). We included the effect of limb darkening, considering a quadratic model with coefficients $[0.425, 0.298]$ (Feliz et al. 2019).

These spectra were fed into `pyechelle`, incorporating realistic airmass distributions to model telluric lines and barycentric radial velocities, with simulated exposures of 2 minutes each. To minimize radial velocity shifts due to instrumental effects and barycentric radial velocity differences, exposures at opposite values of $\theta$ were taken consecutively. While RISTRETTO was not explicitly designed for high-precision radial velocity measurements, it can achieve an instrumental stability of 1 m/s over short timescales such as a few minutes. The resulting 1D spectra from these simulated observations were analyzed using the approach in Bouchy et al. (2001) to compute the radial velocity shift between opposite exposures. The aim was to identify the position angle at which the radial velocity difference between opposite exposures was minimal, corresponding to the orientation of the stellar spin axis. To compute the radial velocity shift we used the following equations (Eq. 3, Bouchy et al. (2001)):

$$\frac{RV_j}{c} = \frac{F_{a,j} - F_{b,j}}{\lambda_j} \cdot \left| \frac{\partial F_{0,j}}{\partial \lambda_j} \right|^{-1}, \tag{15}$$

$$\frac{\sigma_{RV_j}}{c} = \frac{\sqrt{\sigma^2_{F_{a,j}} + \sigma^2_{F_{b,j}}}}{\lambda_j} \cdot \left| \frac{\partial F_{0,j}}{\partial \lambda_j} \right|^{-1}, \tag{16}$$

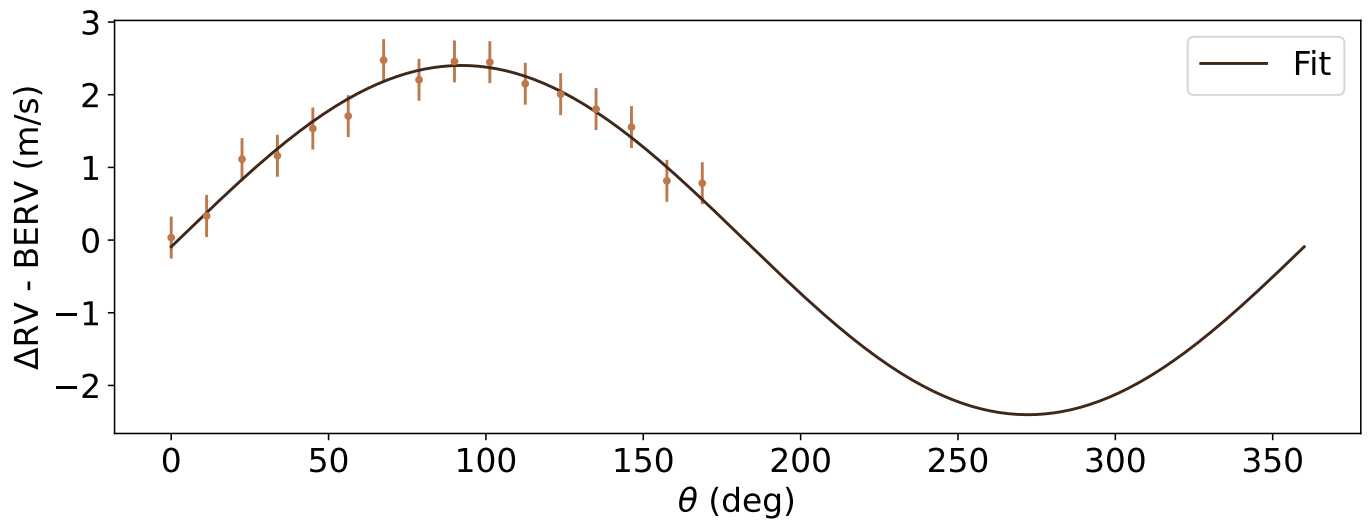

Fig. 22: Measured radial velocity differences between opposite exposures as a function of the position angle of the star within the central spaxel of RISTRETTO. The star is placed at a constant distance of 9 mas from the center.

where $j$ represents the $j^{th}$ pixel in a given order, $F_0$ represents a reference high-S/N spectrum constructed by summing all 32 exposures, and $F_a$ and $F_b$ are spectra of opposite exposures. Finally, $\delta V$ was computed by solving the linear equation 15.

The radial velocity difference curve is shown in Figure 22. It was modeled using the following equation:

$$\Delta RV(\theta) = A \cdot \sin(\theta - \theta_s), \tag{17}$$

where $\theta_s$ is the orientation of the stellar spin axis and $A$ is the RV semi-amplitude of the signal. The latter depends on the stellar equatorial velocity given by $2\pi R_\star / P_{\rm rot}$, the stellar inclination angle, and the fiber coupling curve.

A Markov chain Monte Carlo (MCMC) approach (Foreman-Mackey et al. 2013) was employed to fit this model to the observed data, yielding the posterior estimates for $\theta_s$ and $A$. Figure 22 shows the data and the fitted result (MCMC posteriors are shown in Appendix B). In addition, we performed ten different simulations to account for different noise realizations and test the robustness of our data analysis. The ten fitted $\theta_s$ values are shown in Figure 23. Their mean value and scatter match the true value used in the simulation ($\theta_s = 0$) and estimated error bars. We also tested different orientations other than zero, and the procedure worked equally well.

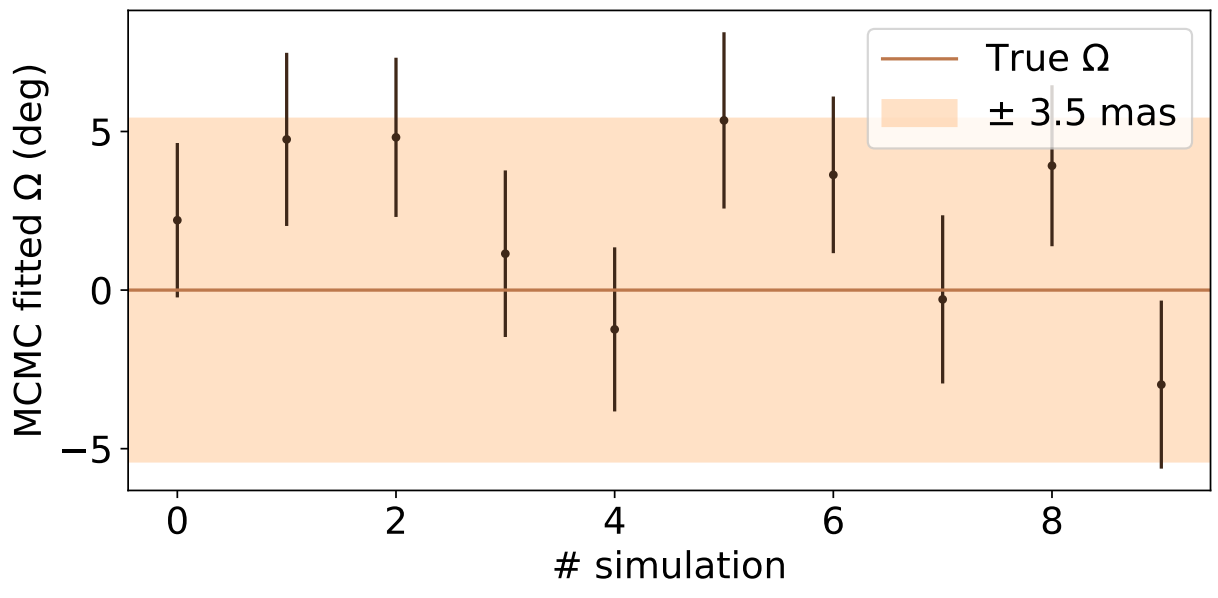

Fig. 23: Estimated $\theta_s$ from the MCMC posteriors for 10 different simulations. Each value represents the median of the posterior, and the errors are computed as the $16^{th}$ and $84^{th}$ percentiles. The horizontal line shows the true $\theta_s$ value used in the simulation. The shaded region illustrates an angular range of $\pm 5.4°$, which corresponds to a tangential offset of 3.5 mas at 2 $\lambda/D$ for the RISTRETTO IFU.

All calculated values of $\theta_s$ fall within $\pm 5.4°$. Assuming the stellar equatorial plane is aligned with the orbital plane of the planet, constraining its orientation to within $\pm 5.4°$ ensures that the planetary position at maximum elongation deviates by no more than 3.5 mas from the off-axis fiber center, corresponding to about 20% of coupling loss. This corresponds to approximately one-third of the constraint used to select the observational epochs. We thus expect that RISTRETTO will be able to determine the stellar spin axis orientation with sufficient accuracy to enable an optimal choice of the IFU position angle, thereby saving up to a factor of 2 in observing time if the planet orbit is aligned with the stellar equator.

We show that the stellar inclination angle could be derived from the measured value of the $A$ term and the known stellar equatorial velocity, angular diameter, and fiber coupling curve. It is, however, unclear to what precision it could be measured, given the uncertainties on the coupling curves. We leave this issue to a future work.

## 7. Discussion and conclusion

### 7.1. Summary

This study demonstrates the realistic potential of the RISTRETTO instrument to detect Proxima b in reflected starlight, marking a critical step forward for the atmospheric characterization of nearby rocky exoplanets. Through detailed end-to-end simulations, we modeled realistic observational conditions, instrument behavior, and data analysis pipelines, showing that Proxima b can be detected in approximately 55 hours of observation, assuming an Earth-like atmosphere, and its mean albedo constrained at the 20% level. The presence of molecular features in the planetary atmosphere is supported by the data: by fitting the "true" high-resolution synthetic albedo spectrum used in the simulations, we can reach a statistically significant detection in about 85 hours of observations. Simplified molecular models also support this sensitivity to $O_2$, suggesting that $H_2O$ could be detectable with exposure times exceeding 100 hours. A variety of planetary atmosphere models, such as an "agnostic" chromatic, broadband albedo spectrum, can be fitted to the data to explore the existence and shape of potential molecular bands and scattering processes.

Furthermore, a novel technique that makes use of the unique light-coupling properties of single-mode fibers allows us to determine the sky orientation of the stellar spin axis in less than two hours of observing time, providing critical information to predict the apparent planetary orbit and optimize spaxel alignment.

These results underline the capabilities of RISTRETTO not only as a pathfinder for next-generation instruments on extremely large telescopes such as ELT-ANDES and ELT-PCS, but also as a standalone scientific instrument capable of pioneering reflected-light spectroscopy of temperate terrestrial planets.

### 7.2. Future improvements and directions

Some key points that could be investigated in future studies are given below.

1. In the analyses performed here, we chose short exposures bracketing each long exposure as our template to remove the stellar and telluric lines from the long exposures. Instead, we could consider using the stellar+telluric spectra recorded during the long exposure itself in the other spaxels; for example, the central one. This approach would provide a more accurate model for telluric absorption, barycentric correction, and potential stellar activity. The downside is that this template would not have as high S/N as the short exposures and would be measured in different spaxels, which will introduce systematic effects at the level of several percent due to the varying spectrograph PSF between spaxels. Therefore, such a method would require an in-depth characterization of how the PSF affects the spectra. We leave this aspect to a future work.
2. It is challenging to produce model planetary spectra that account for the detailed molecular line shapes to which high-resolution spectroscopy is sensitive. To reduce degeneracies and fully exploit the high-resolution spectral content of the data, a promising approach for future analysis could involve constructing a library of GCM-based spectra, covering a range of realistic atmospheric scenarios. By fitting these spectra directly to the data, the method would effectively shrink the free parameter space associated with atmospheric properties. In the Bayesian framework, this approach reduces the integral over these parameters to a discrete (or near-delta function) evaluation, streamlining the comparison while still extracting meaningful insights into the planet atmospheric composition.
3. In the simulations performed as part of this study, we assumed a planetary spectrum observed at a fixed orbital phase of 90 degrees. More sophisticated simulations and models could probe how detections might improve or worsen using a range of orbital phases. Specifically, as the planet approaches superior conjunction, its reflected flux increases substantially, resulting in a stronger signal. However, at the same time, its angular separation from the star decreases, decreasing the planet coupling into the off-axis spaxels. These two competing effects could be included in the modeling to probe the planetary phase curve; for instance, by including a simple phase function in the model (Lambertian or locally linear around quadrature).
4. Regarding the instrument simulator, further refinements could focus on a more sophisticated detector model, incorporating factors such as bias residuals, cosmic ray hits, hot pixels, remanence, and similar effects.
5. Regarding exposure selection, a possible improvement would be to retain all frames and perform a weighted fit so that higher-quality exposures contribute more. In principle, the weights could be derived from the coupling-function ratios defined in Section 4.3.1. However, the third ratio contains $\rho_i^p(\mathbf{r})$, which depends on the (a priori) unknown planet position within the off-axis spaxel at each epoch. Using such weights would thus implicitly assume an orbital geometry. To avoid imposing strong priors on the orbit, we adopted a conservative approach and use uniform weights. A practical alternative would be an iterative scheme: (i) perform an initial unweighted fit to detect the planet and estimate its orbit; (ii) recompute exposure weights from the inferred geometry; and (iii) refit to improve the parameter constraints. Implementing and validating this procedure is beyond the scope of the present work.

Looking forward, this work provides a foundational methodology for detecting and characterizing exoplanets in reflected light around nearby stars with ground-based telescopes. The RISTRETTO instrument concept and the data analysis techniques developed here pave the way for next-generation instruments on extremely large telescopes such as ELT-ANDES and ELT-PCS.

### 7.3. Other targets

RISTRETTO is designed not only to focus on Proxima b but aims at pioneering reflected-light spectroscopy for a range of other known exoplanets, including gas giants, Neptunes and super-Earths. Before attempting to characterize Proxima b, RISTRETTO will first target "easier" exoplanets. These include the cold gas giants GJ 876 b & c, which can be explored in just 2-3 hours of observing time. Other possible targets are the warm Saturn HD 3651 b, the Neptune-mass planets HD 192310 b, HD 102365 b and 61 Vir d, and the warm super-Earth GJ 887 c which can be probed in a single to a few nights of observations (Lovis et al. 2024). Thus RISTRETTO is by no means a "single-target" instrument but truly has the capability to open a new parameter space in exoplanet studies.

## Appendix A: Barycentric radial velocities during the observational epochs

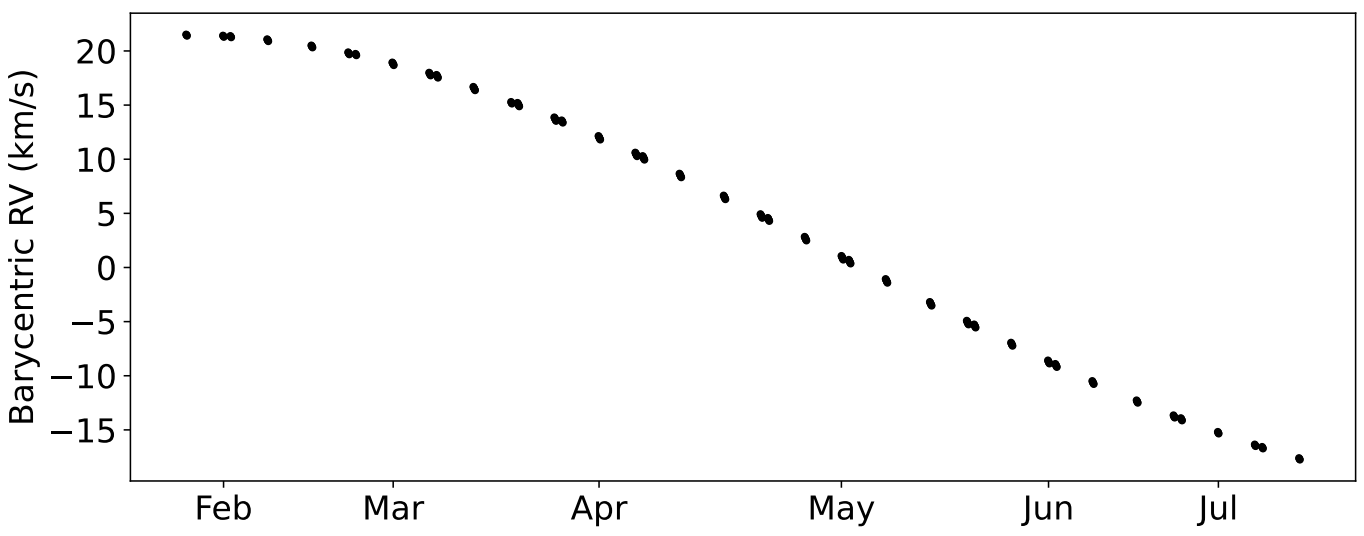

Fig. A.1: Barycentric correction during the observational epochs

## Appendix B: Star spin angle MCMC example result

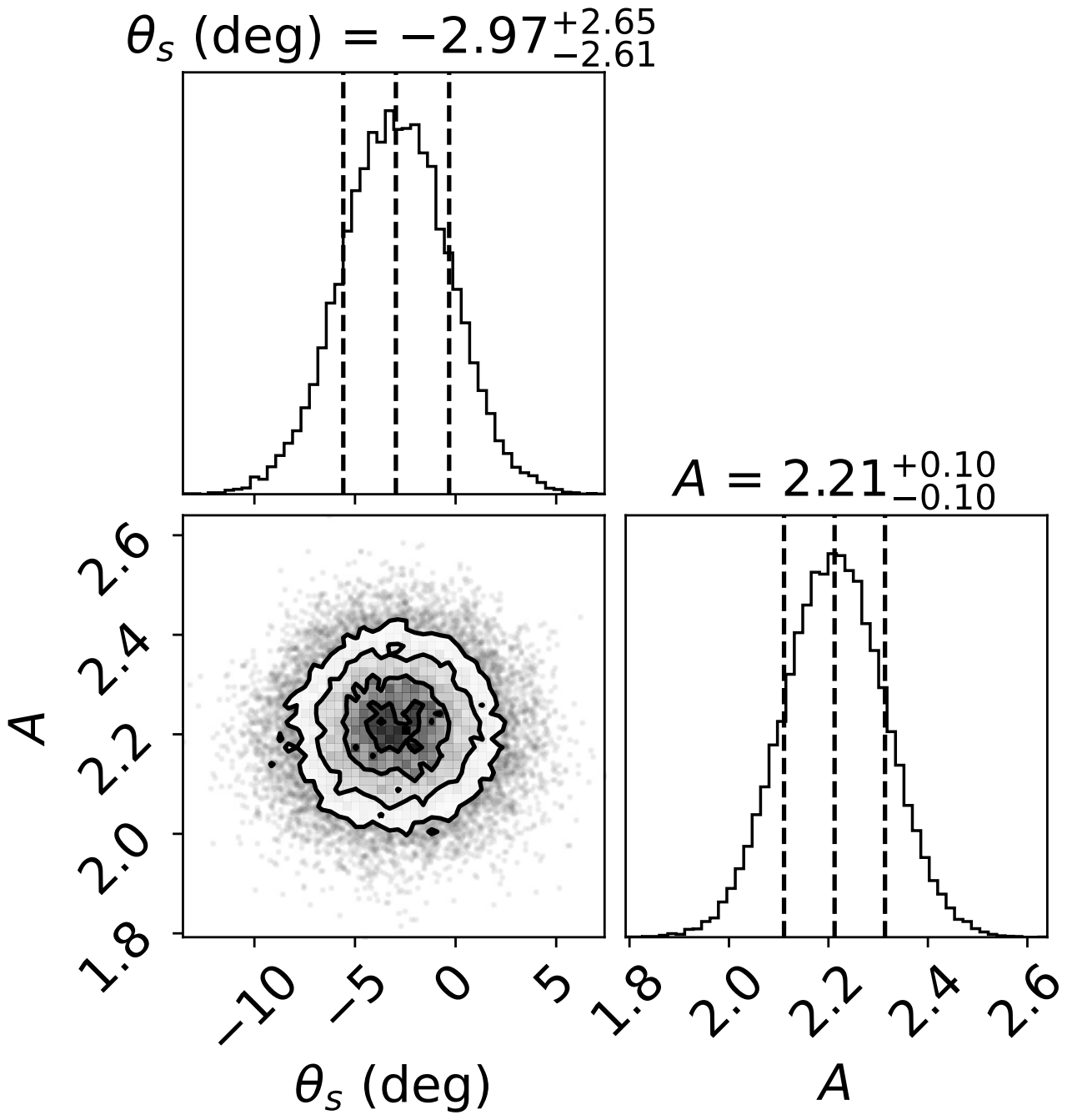

Fig. B.1: MCMC corner plot for one simulation showing the posterior distributions for the orientation of the stellar spin axis $\theta_s$ and RV semi-amplitude $A$.

## Appendix C: Posterior distribution results for the molecular absorption model

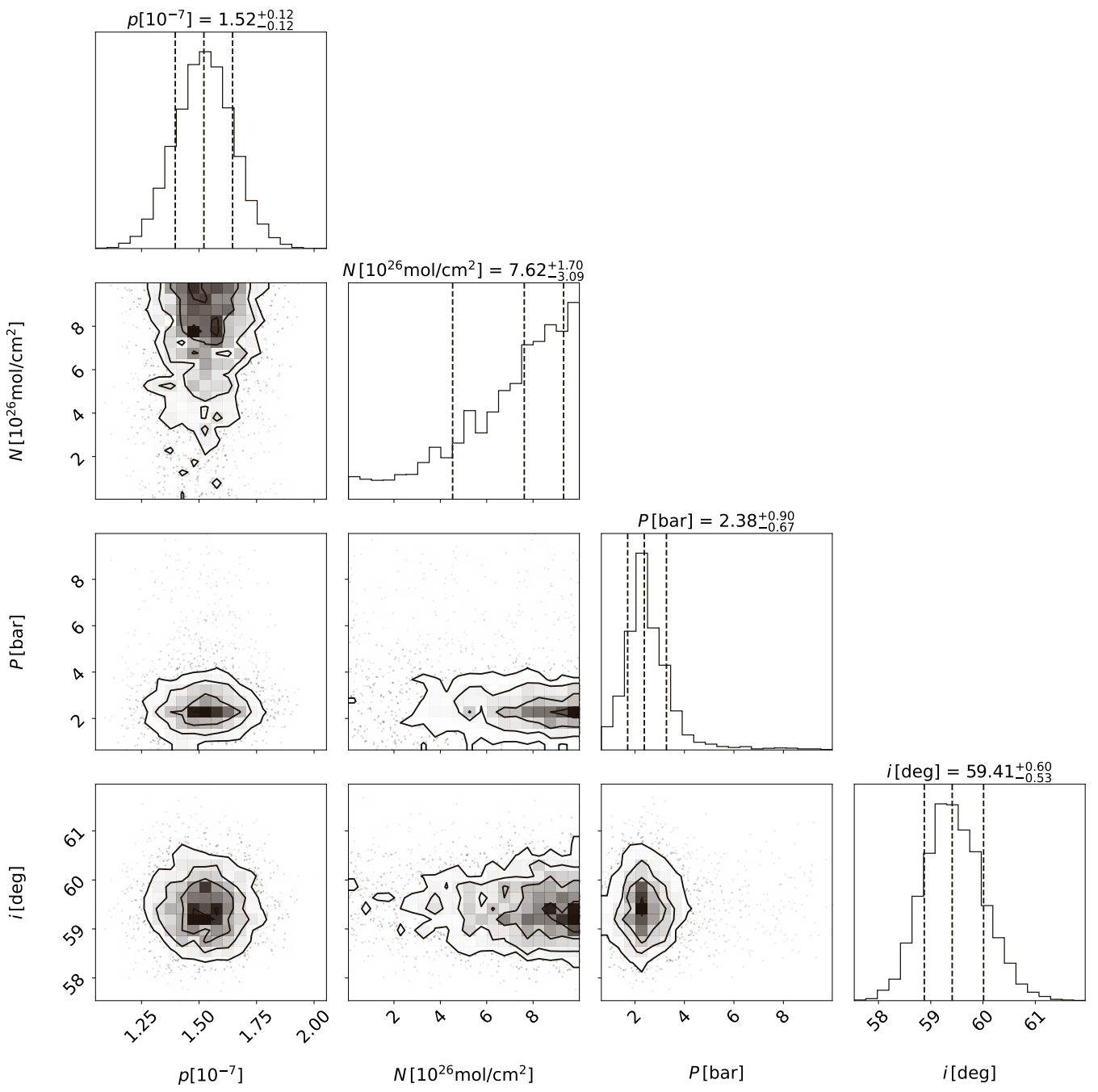

Fig. C.1: Posterior distribution for the $O_2$ molecular model at a fixed temperature of 250 K.

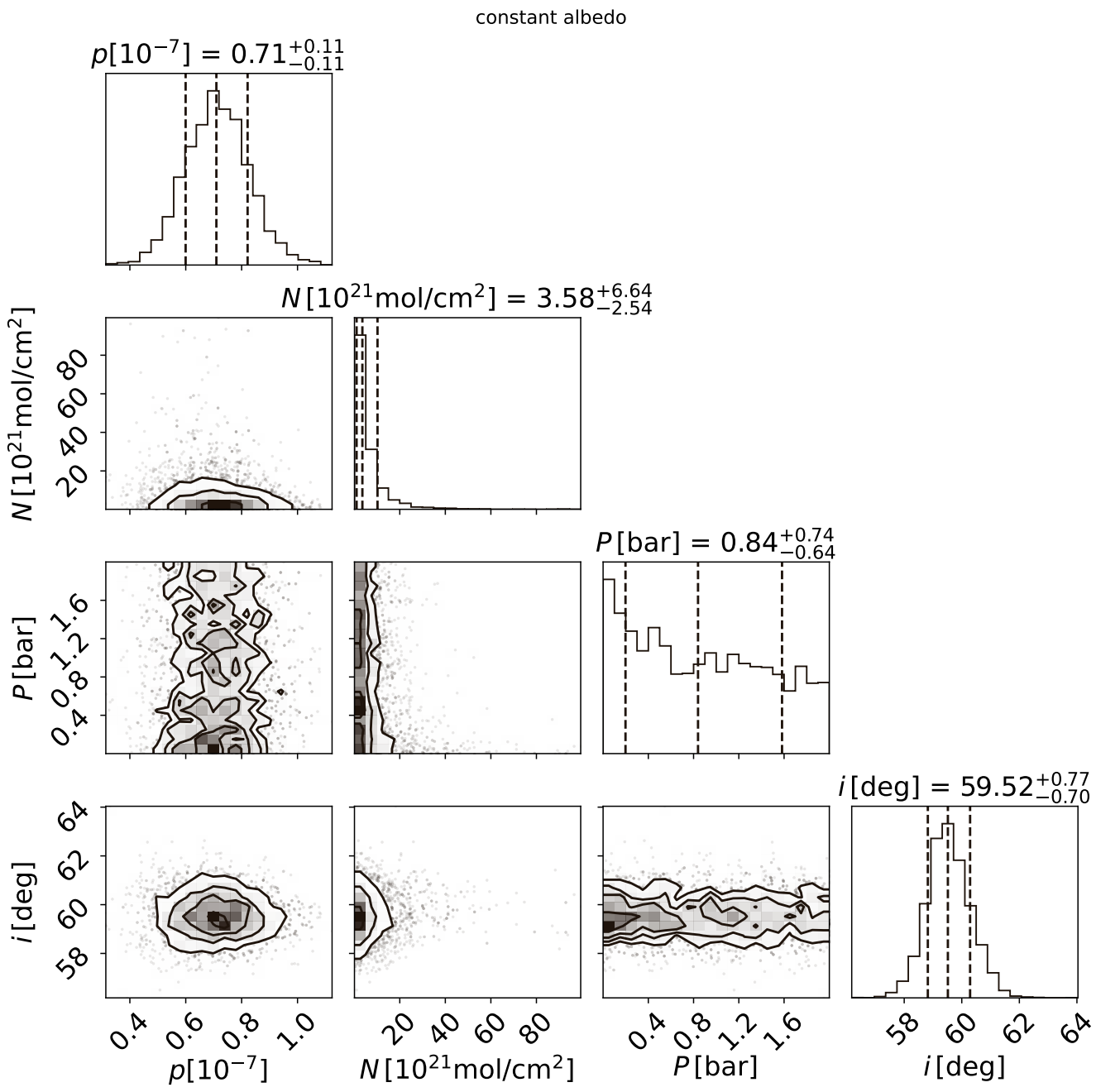

Fig. C.2: Posterior distribution for the $H_2O$ molecular model at a fixed temperature of 250 K.